\documentclass[twocolumn, amsmath, amsfonts, article]{revtex4}
\pdfoutput=1
\usepackage{graphicx}
\usepackage{subfigure}
\usepackage{threeparttable}
\usepackage{array}
\usepackage{hhline}
\usepackage{enumitem}
\newcolumntype{P}[1]{>{\centering\arraybackslash}p{#1}}
\newcolumntype{Q}[1]{>{\raggedright\arraybackslash}p{#1}}
\begin{document}
\title{Comprehensive numerical modeling of filamentary RRAM devices including voltage ramp-rate and cycle-to-cycle variations}
\author{Dipesh Niraula}\email{dipesh.niraula@rockets.utodeo.edu}\author{Victor Karpov}\email{victor.karpov@utoledo.edu}\affiliation{Department of Physics and Astronomy, University of Toledo, Toledo, OH 43606, USA}
\begin{abstract}
The equilibrium ON and OFF states of resistive random access memory (RRAM) are due to formation and destruction of a conducting filament. The laws of thermodynamics dictate that these states correspond to the minimum of free energy. Here, we develop a numerical model that, through the minimization of free energy at a given voltage, determines the filament parameters and thus the electric current. Overall, it simulates the current-voltage (I-V) characteristics of RRAM. The model describes mutual transformations of RRAM states through SET (ON$\rightarrow$OFF) and RESET (OFF$\rightarrow$ON) processes. From the modeling perspectives, these states and processes constitute four programming modules constructed here in COMSOL Multiphysics software tackling the electrodynamic and heat transfer equations and yielding RRAM energy and I-V. Our modeling uniquely reproduces the observed I-V varying with voltage ramp-rates. That is achieved by accounting for the ramp-rate dependent activation energy of conduction. The underlying mechanism is due to the deformation interaction caused by the double well atomic potentials universally present in amorphous materials and having exponentially broad distribution of relaxation times. As another unique feature, our modeling reproduces the observed cycle-to-cycle variations of RRAM parameters attributed to the lack of self-averaging in small ensembles of double well potentials and electronic states in geometrically small (nano-size) RRAM structures.
\end{abstract}

\maketitle

\section{Introduction}
\subsection{General}\label{sec:gen}
\begin{figure}[!ht]
  \centering
  $\vcenter{\hbox{\includegraphics[width=0.19\textwidth]{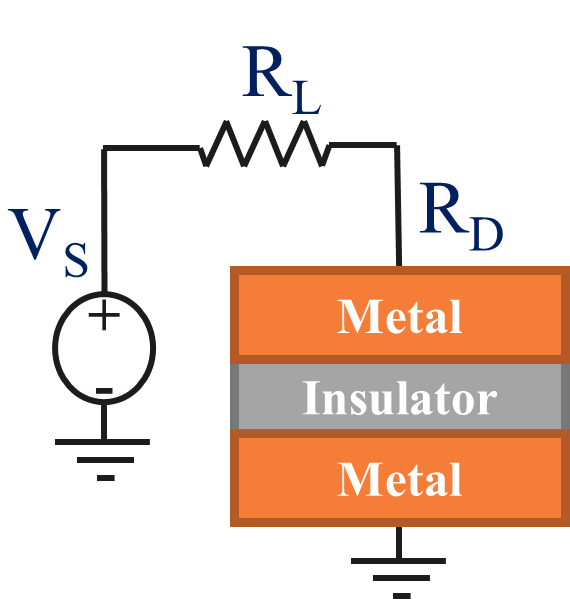}}}$$\vcenter{\hbox{\includegraphics[width=0.3\textwidth]{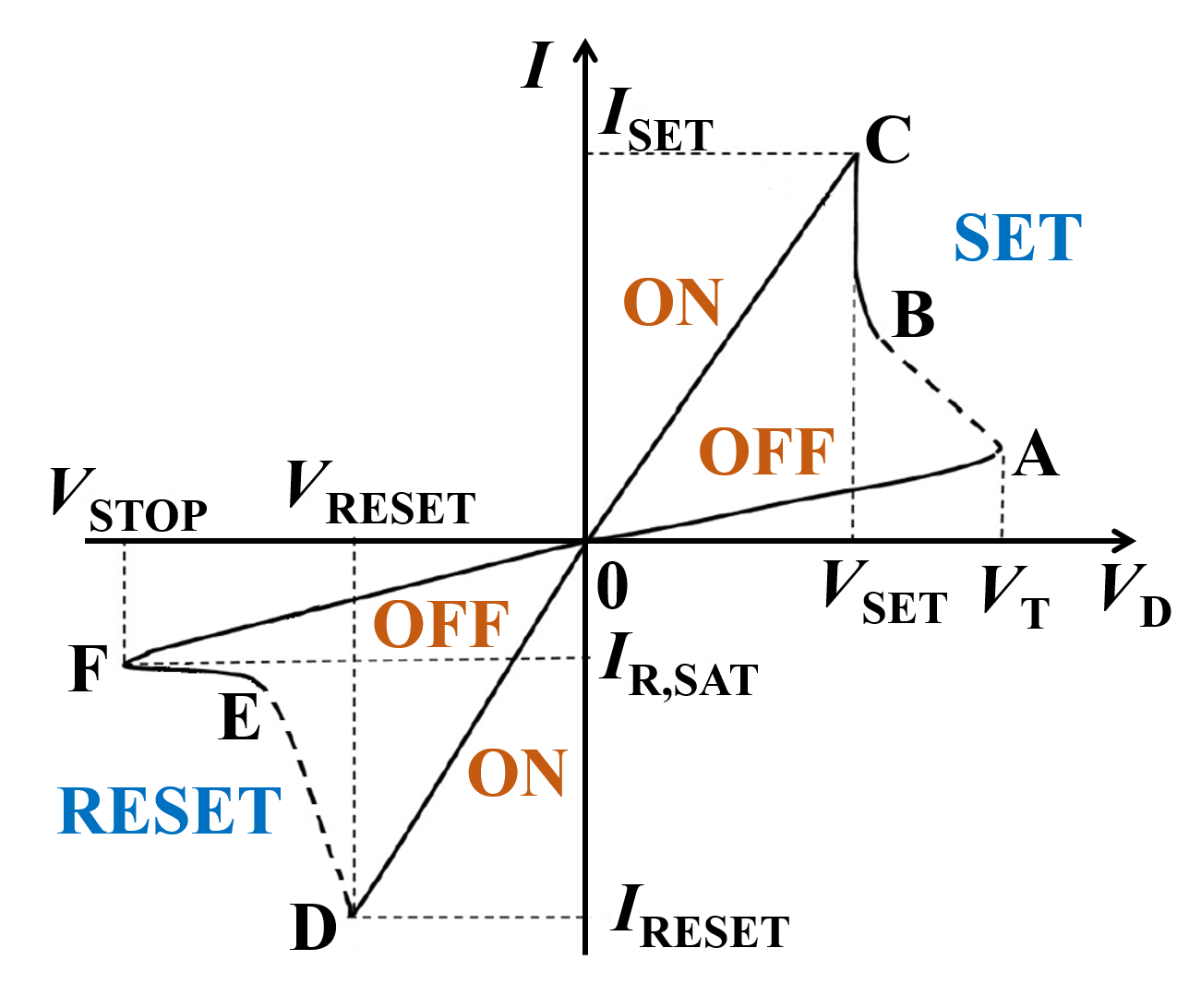}}}$
  $\vcenter{\hbox{\includegraphics[width=0.19\textwidth]{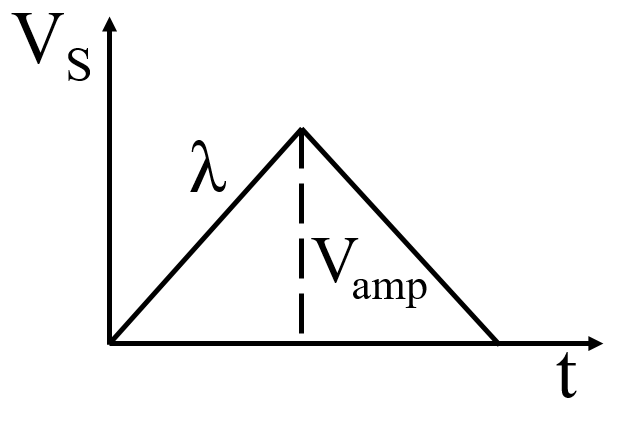}}}$$\vcenter{\hbox{\includegraphics[width=0.3\textwidth]{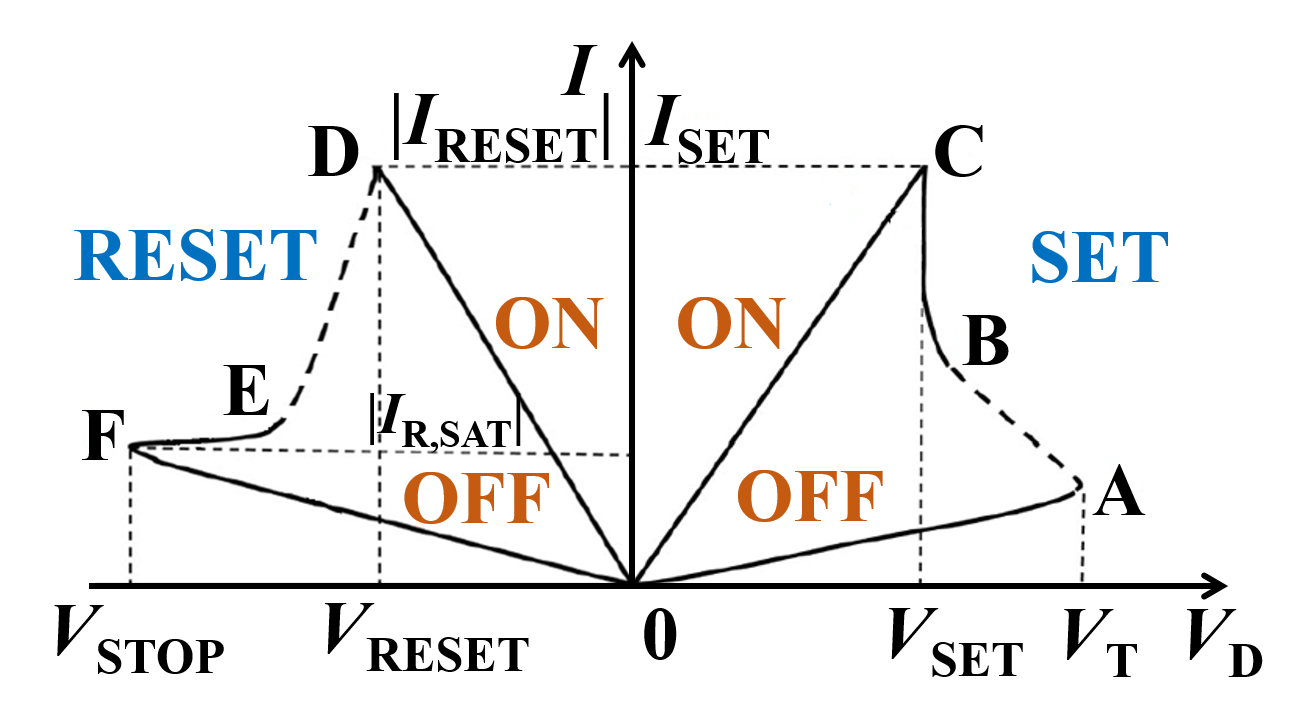}}}$
  \caption{Top left: Schematic of the circuitry where input voltage source $V_S$, load resistor $R_L$ and a simple RRAM device $R_D$ are connected in series. Bottom left: plot of input voltage pulse where $\lambda$ is the voltage ramp-rate and $V_\textrm{amp}$ is the amplitude. Top right: Representation of a typical I-V characteristics of RRAM \cite{Fantini2012,Wouters2012,Chen2012,Ambrogio2014}. $V_D$ is the voltage drop across RRAM device. The SET current $I_\textrm{SET}$ is the maximum current limited by the $V_\textrm{amp}$. Bottom right: Customary representation of the I-V with absolute valued current axis. The dashed domain represents very fast nucleation processes.}\label{Fig:RRAM_IV}
\end{figure}

RRAM is a metal-insulator-metal multilayered device sketched in Fig. \ref{Fig:RRAM_IV} (top left). It operates via resistive switching between the insulating and conducting states in response to electric bias. The switching is due to formation or destruction of a conductive filament. The corresponding conducting (ON) and insulating (OFF) states represent the logical `1' and `0'. Since RRAM is energetically economical and can be densely packed in a crossbar array, it became a promising non-volatile memory actively researched in the last decade.

The recent thermodynamic approach \cite{Karpov2016,Karpov2017} describes RRAM switching in terms of phase transformations. As a next step dictated by technology needs, here, we develop a comprehensive quantitative thermodynamic based numerical model of RRAM, briefly introduced earlier in Ref. \cite{Niraula12017}.
\begin{figure*}[!ht]

\centering
\includegraphics[width=0.55\textwidth]{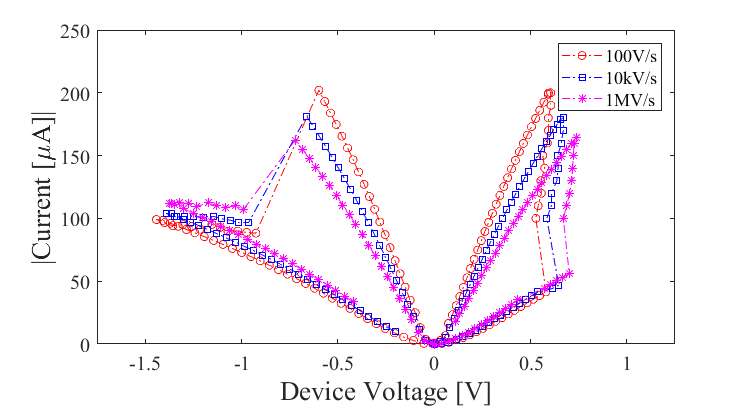}\label{Fig:CompleteIV} \includegraphics[width=0.25\textwidth]{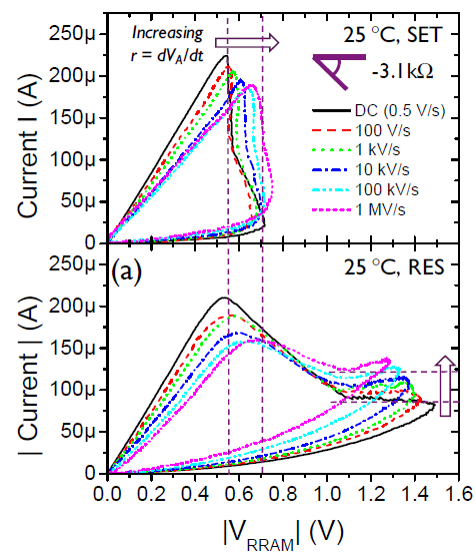}
\includegraphics[width=0.55\textwidth]{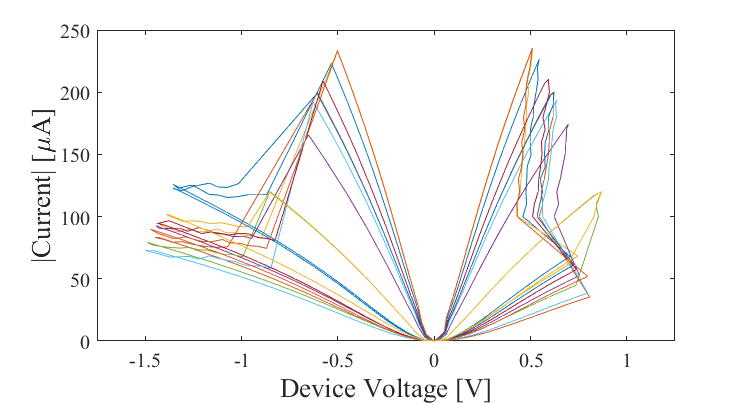}\label{Fig:rand1}\includegraphics[width=0.25\textwidth]{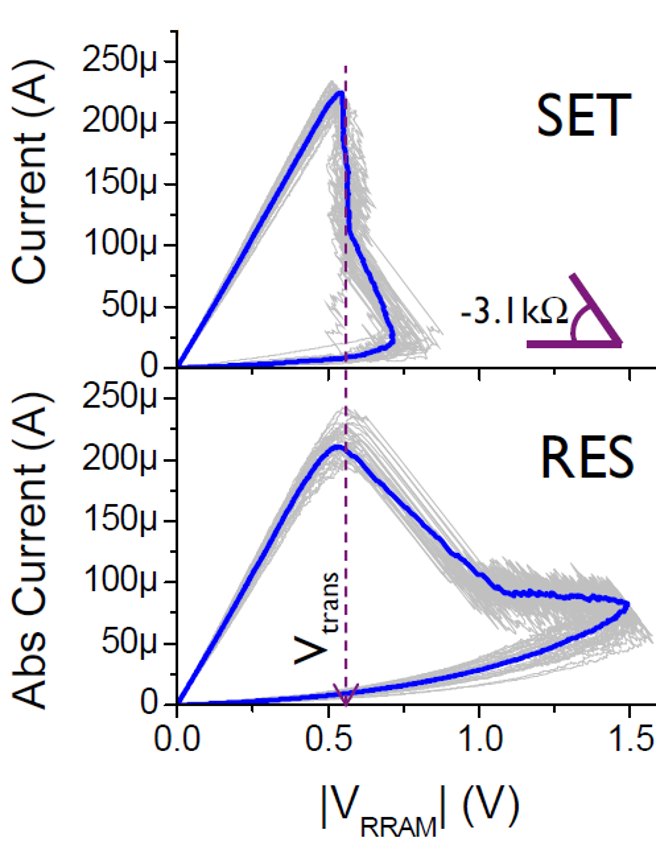}
\caption{Comparison of simulated (left) and experimental (right) (\textcopyright 2012 IEEE. Reprinted with permission from \cite{Fantini2012}. ) I-V for identical device. Top figures represent the applied voltage ramp-rate dependent I-V and bottom figures represent cycle-to-cycle variation in I-V. In right figures, $V_\textrm{RRAM}$ is the device voltage, $V_\textrm{trans}$ is the SET voltage, 3.1 k\textrm{$\Omega$} is the load resistor, and 25$^\circ$C is the room temperature. SET and RES refers to SET and RESET process respectively. The arrows in the top and bottom picture represent the increasing SET voltage and saturation current with faster pulse. The bottom right picture shows the I-V for 50 cycles.}\label{Fig:IV}
\end{figure*}

We model RRAM in COMSOL multiphysics software that solves partial differential equations via finite element method. The electrodynamic and heat transfer equations were solved for a given device structure utilizing the lump material parameters.

A sketch of the typical I-V in Fig. \ref{Fig:RRAM_IV} (right), consists of several domains related to the device states and processes. The domains C-0-D and A-0-F correspond to the ON and OFF states remaining intact under voltage ramping. On the contrary, A-B-C and D-E-F reflect structural and electric current variations in response to the source voltage ramping; they represent the SET and RESET processes. Their  characteristic features are voltage snapback A-B followed by the vertical domain B-C for SET, and voltage snapforward D-E followed by the horizontal domain E-F for RESET.

\subsection{Numerical experimenting}\label{sec:numex}The development of this modeling program evolved through a number of numerical experiments that, upon comparison with the available data, allowed us to make definite conclusions about the physics underlying RRAM operations; that would not be possible otherwise due to a large number of conceivable alternative mechanisms. As an instructive example, we point here at the simplest model of diffusive conduction and heat transport sufficient to generate the basic shape of I-V in Fig. \ref{Fig:RRAM_IV} (right). However, assuming such a transport, further quantitative verifications revealed an `unphysically' high local temperature ($\sim 4000$ K) in a small dielectric region of a ruptured filament. That problem was solved by experimenting with various heat transport mechanisms and recognising the ballistic non-local Joule heat dissipation by the electrons. That numerical experiment result defines the mechanism of electric and heat transport in nano-sized devices. Other important `numerical experiment' pertains to the recognition of the double well potentials (DWP) inherently present in non-crystalline structures. That made it possible to describe the observed I-V variations with voltage ramp-rates and between switching cycles.

\subsection{Outline}\label{sec:outl}
Here we present a simplified ``low-resolution" guide through subsequent consideration, some parts of which present inevitable distracting details. We start with pointing at Fig. \ref{Fig:IV} that plays the role of this research showcase. It illustrates how our modeling uniquely reproduces the observed RRAM characteristics including I-V variations with voltage ramp-rate and cycle-to-cycle. More in details, the paper is organized as follows.
\begin{enumerate}
  \item {\bf Physics} - In Sec. \ref{Sec:Physics}, we present the physics underlying RRAM operations. The formation of conductive filament starts with the field induced nucleation of a needle-shaped embryo. As its length increases, it shunts the device and then grows radially, during which process the resistance becomes reciprocal of current corresponding to domain B-C . The filament has a finite capacitance and thus accumulates the charges. The charged filament creates radial polarization that freezes upon the completion of SET process and remains intact in the ON state. The polarization-charge interaction decreases the filament energy. Reversing bias polarity charges the filament oppositely making the earlier created polarization energetically unfavorable. When the reversed bias gets sufficiently large in absolute value, the polarization-charge interaction increases the system energy enough to break the filament introducing the insulating gap in its structure, which constitutes RESET.

  \item {\bf Modeling and Logic} - In Sec. \ref{Sec:Num}, we present details of the numerical modeling.  We implement the mechanism of field induced nucleation \cite{Karpov2008} for SET and that of the gap nucleation for RESET processes. Following nucleation, the filament and the gap growth stages are described as dissipative processes where the temporal derivatives of the filament radius and gap width are proportional to the corresponding thermodynamic forces. The equilibrium states are achieved when those forces vanish, i. e. when the free energy is a minimum. Along these lines, minimization of free energy under given bias generates the filament parameters, hence, resistance and electric current, predicting I-V shape.

      We have successfully simulated I-V with ramp-rate dependence and cycle-to-cycle variation as shown in Fig. \ref{Fig:IV}. We devised a program simulating actual device operations. The program consists of control, six modules that represent the device states and switching processes, along with switching conditions. The modules are constructed in COMSOL, while the program is written in MATLAB, which switches the modules when the switching criteria are fulfilled.

  \item {\bf Role of Disorder} - In Sec. \ref{SEC:Disorder}, we introduce configurational characteristic of non-crystalline materials, such as metal oxides used as the insulator layer in RRAM. In such materials, a certain fraction of atoms or groups of atoms retain their mobility having two equilibrium positions described as double well potentials (DWP). The transition rate between the two minima is determined by the height of their activation barrier. The applied bias can activate DWP with relaxation times shorter than the pulse length; hence, the ramp-rate dependence. Because of the very small volume of a device, the number of DWPs in it is not large enough to provide the statistical self-averaging: different implementations of nominally identical structures will have different ensembles of DWPs, which is responsible for the cycle-to-cycle variations in our numerical modeling.

  \item {\bf Details of COMSOL} - In Appendix \ref{Ap:DevModel}, we present more specific details of modeling in COMSOL, such as algorithm, device geometry and dimension, utilized differential equations and boundary conditions, minimization procedure, and various parameters, aimed at allowing the reader to replicate our modeling.
\end{enumerate}

\section{Physics of Device Operation}\label{Sec:Physics}

\begin{figure}
  \centering
  \includegraphics[width=0.35\textwidth]{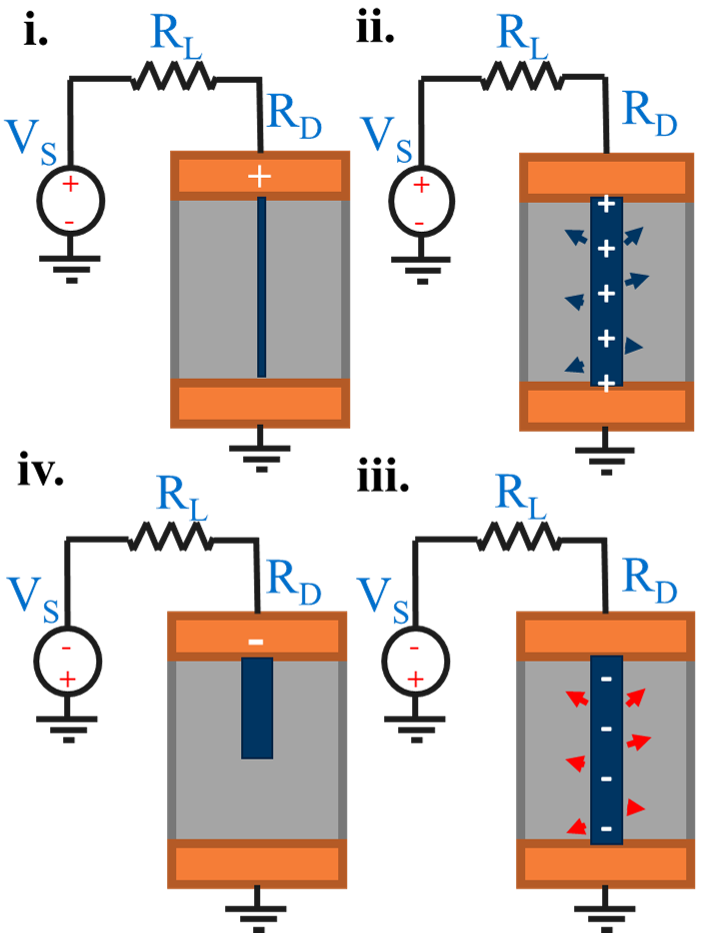}
  \caption{Formation and dissolution of the filament in a generic metal-insulator-metal multilayered RRAM device. The arrows represent polarization of the insulating host near filament; blue (ii) and red (iii) color denotes energetically favorable and unfavorable states, respectively. Note: the pictures are placed in a cyclic order corresponding the SET and RESET switching cycle.}\label{Fig:Thermo}
\end{figure}
\begin{figure}
  \centering
  \includegraphics[width=0.3\textwidth]{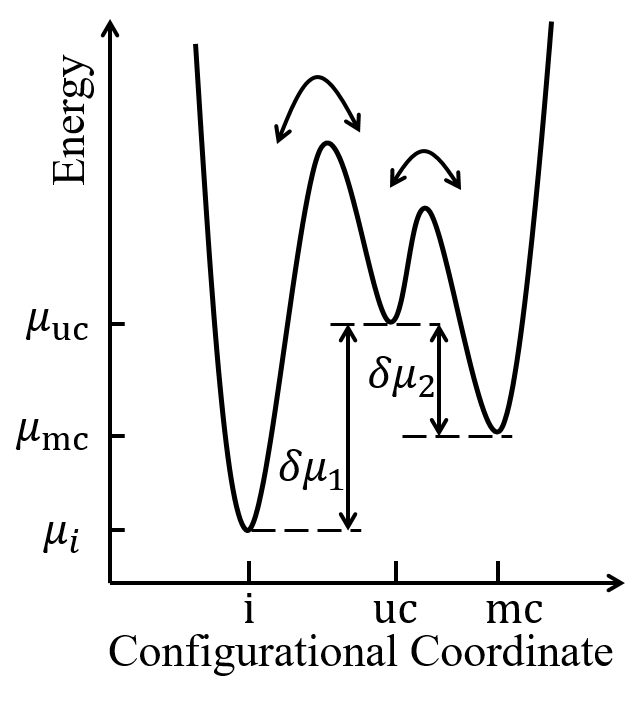}
  \caption{1D presentation of the system chemical potentials in unspecified configurational coordinates
showing three distinct minima corresponding to the insulating (i), unstable conductive (uc), and metastable conductive (mc) phases and their related barriers. Arrows represent transformations between the mc and uc and the uc and i phases where the energy barriers are relatively low.}\label{Fig:phase}
\end{figure}

Here we briefly recall the thermodynamic theory of RRAM. \cite{Karpov2017} Three concepts form its basis. \\
(1) The concept of radial electric field polarizing the surroundings: Because the filament has a finite capacitance, it acquires the linear electric charge corresponding to bias polarity. That charge creates the radial field that polarizes the surrounding material by aligning the existing electric dipoles and/or displacing the charged defects (vacancies, etc.). The interaction between the charge and its created polarization provides a significant contribution to the system free energy. Abruptly switching the bias polarity while retaining the polarization will make the system energetically unfavorable and conducive of filament rupture via the nucleation of a narrow insulating gap. Such is the mechanism of RESET illustrated in Fig. \ref{Fig:Thermo}.\\
(2) The concept of at least three different structural phases involved: must be accepted on empirical grounds stating that both the SET and RESET processes are endothermic. Indeed, should the system involve only two phases, such as corresponding to the ON and OFF states, then having one phase transition endothermic would automatically make the reverse transition exothermic. Therefore, we have to assume at least one more phase that is higher in energy than the former two. Furthermore, also on empirical grounds, because the conductive filament can be eliminated by annealing, one has to accept that, of those three phases, the minimum energy must be assigned to the dielectric. In summary, the three phases involved in phases transformations are insulating (i), unstable conducting (uc), and metastable conducting (mc) as shown in Fig. \ref{Fig:phase}. The long-lived conductive filament is composed of metastable conducting phase which transforms to the stable insulating phase during dissolution via the unstable conductive phase and vice-versa. For our numerical modeling, the differences in chemical potentials between the three phases provide sufficient description; these differences are the model adjustable paraemters.\\
(3) The concept of quasi-equilibrium allowing the thermodynamic (rather than more involved kinetic) description: The quasi-equilibrium is due to the very short thermalization time $\tau _T$ related to the device smallness. Indeed, $\tau _T\approx h^2/\kappa _d$ is shorter than 1 ps for the filament height $h\sim5$nm and thermal diffusivity $\kappa _d \sim 0.1-1$ cm$^2$/s, which is much shorter than the device operation times. Because the system is approximately  in thermal equilibrium throughout switching, the application of thermodynamics is validated.\\

With the above concepts in mind, the sequence of device operations can be described as follows, starting from OFF state. Due to the thinness of insulating layer, a moderate switching voltage $V_T$ creates a strong electric field (up to $\sim 10^9$ V/m) increasing the electrostatic energy. In response, the insulating layer locally nucleates into the conducting phase forming a cylindrical embryo as shown in Fig. \ref{Fig:Thermo}(i) [field induced nucleation \cite{Karpov2008}] corresponding to point A in Fig. \ref{Fig:RRAM_IV} (right). The embryo focuses the field on its tip and grows longitudinally even faster forming a filament and shunting the device. The device voltage drops (snapback), and the current flow increases (I-V domain A-B) due to the sudden decrease in device resistance.

Upon shunting, further raising the power source voltage only slightly increases the device voltage because the load resistance dominates that of device ($R_L\gg R_D$). The current increase is due to filament growing radially and decreasing its resistance. This corresponds to the vertical I-V domain B-C where the current grows at a fixed device voltage, $V_\textrm{SET}$. The full-grown filament radius is set by the highest allowed (compliance) current ($I_\textrm{SET}$ in Fig. \ref{Fig:RRAM_IV} (right)).

When the power source polarity is reversed the filament becomes energetically unfavorable as explained in item (1) above. It dissolves when the temperature raises enough to unfreeze the atomic mobility. The corresponding Joule heat must be then equal to that under SET voltage; hence the condition $V_\textrm{SET}=|V_\textrm{RESET}|$ (point D in Fig. \ref{Fig:RRAM_IV} (right)). The dissolution is described as the nucleation of the gap where the voltage snaps forward and electric current drops instantaneously due to the abrupt increase in the device resistance (I-V domain D-E). After switching to OFF state, the device resistance dominates that of the load ($R_D\gg R_L$). Increasing the absolute value of source voltage increases the gap width and the currents saturates represented by $I_\textrm{R,SAT}$.

\section{Numerical Model}\label{Sec:Num}
The primary goal of the numerical modeling is to reproduce the observed I-V characteristics of RRAM. Specifically, the I-V is generated with a pulse voltage source and a load resistor ($R_L$) in series to RRAM, in contrast to a more common method where a transistor is used instead of the load resistor.  The advantage of the former is that one can extract the device I-V by subtracting the voltage drop across the load resistor from the source voltage, $V_D= V_{S}-IR_L$. The pulse amplitude automatically puts a limit to the current through the device.

The switching part of numerical model is described as a phase transition initiated by nucleation which is a stochastic process. Modeling that type of stochastic processes goes beyond the current scope. Our model treats nucleation as a deterministic process using analytical solutions for threshold voltage and RESET voltage.

The switching is governed by the thermodynamics, which includes the electrostatic, thermal, and chemical free energy contributions whereas I-V characteristics of the device states depend on the device electric field and temperature distribution. The field and temperature distributions are obtained by solving coupled heat-electromagnetic partial differential equations with COMSOL. Different I-V domains are solved separately and a program was developed to combine the corresponding I-V branches together. The program is written in MATLAB programming package which communicates with COMSOL via LiveLink. The details are presented in Appendix \ref{Ap:DevModel}.

\begin{figure}
  \centering
  \includegraphics[width=0.45\textwidth]{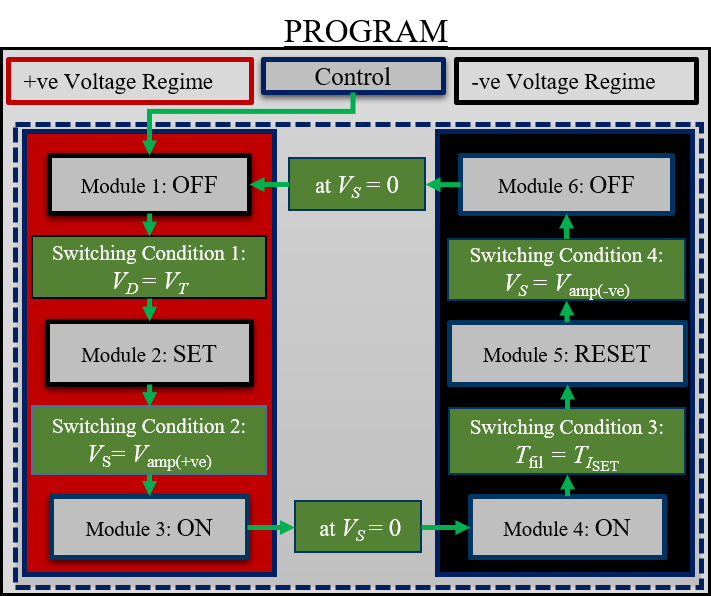}
  \caption{A flowchart of the program that simulates RRAM I-V characteristics.}\label{Fig:program}
\end{figure}
\subsection{Program}
The program sketched in Fig. \ref{Fig:program} simulates complete I-V characteristics of RRAM. It consist of the control, six modules, and four switching conditions. The control is the program's `brain' switching the modules when the corresponding conditions are satisfied. The modules refer to different stationary or dynamic states of the COMSOL modeled RRAM. In particular, the OFF state and RESET process correspond to RRAM with a partial (disrupted) filament as shown in Fig. \ref{fig:dev2018} (bottom). Furthermore, the modules can be naturally divided into positive and negative polarity regimes representative of the actual switching sequence. Table \ref{Tab:program} lists all the I-V domains of Fig. \ref{Fig:RRAM_IV} (right) and their corresponding modules along with the switching conditions. They are explained more in details next.
\begin{table}\footnotesize
  \centering
  \caption{Program and its I-V counterpart (Fig. \ref{Fig:RRAM_IV} (right)).}\label{Tab:program}
  \begin{tabular}{c|c}
    \hline
    \hline
    Program & I-V domain \\
    \hline
    \multicolumn{2}{c}{+ve Voltage Regime}\\
    \hline
    Module 1: OFF & 0-A\\
    Switching Condition 1: $V_D = V_T$ & A \\
    Module 2: SET & B-C\\
    Switching Condition 2: $V_S = V_\textrm{amp(+ve)}$& C \\
    Module 3: ON & C-0 \\
    \hline
     \multicolumn{2}{c}{-ve Voltage Regime}\\
    \hline
    Module 4: ON & 0-D \\
    Switching Condition 3: $T_\textrm{fil}=T_{I_\textrm{SET}}$  & D \\
    Module 5: RESET & E-F \\
    Switching Condition 4: $V_S = V_\textrm{amp(-ve)}$ & F \\
    Module 6: OFF & F-0 \\
    \hline
    \hline
  \end{tabular}
  \end{table}
\subsubsection{{\bf Module 1: OFF}}\label{Sec:M1}
The simulation starts with OFF module dealing with a partial filament whose integrity is compromised by an insulating gap as shown in Fig. \ref{fig:dev2018} (bottom). OFF module corresponds to I-V domain 0-A where the source voltage pulse is ascending in time. The conduction mechanism is critical here. Once it is chosen, the OFF state I-V curve can be generated.

\begin{figure}
  \centering
  \includegraphics[width=0.3\textwidth]{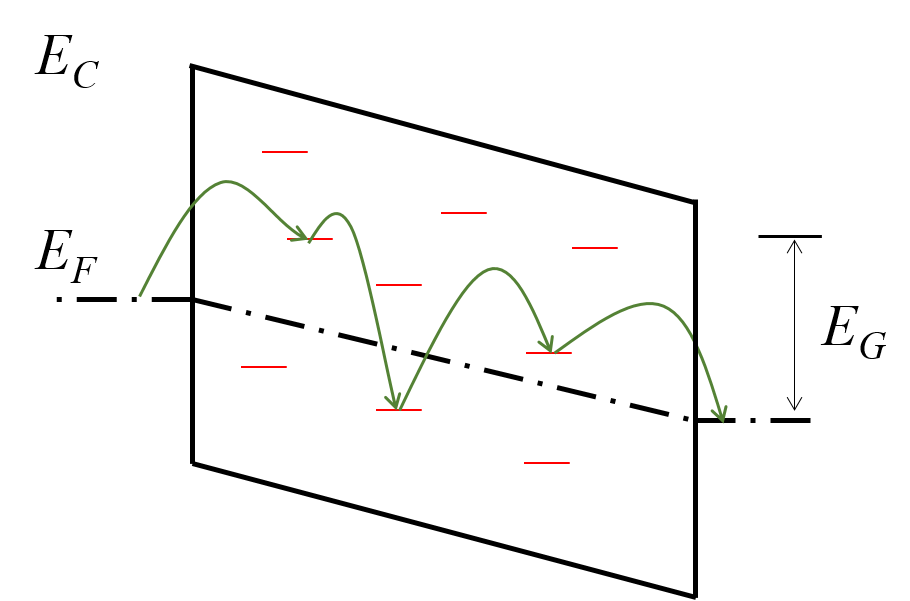}
  \caption{Electron hopping through the optimum path made up of insulator defect sites. $E_F$ is the fermi level and $E_C$ is the mobility edge of the conduction band in an amorphous material.}\label{Fig:hop}
\end{figure}
{\bf Electric Conduction:} The conduction mechanism of filamentary RRAM remains poorly understood. \cite{Lim2015} Various models, such as Poole-Frenkel \cite{Long2012, Mahapatra2015, Walczyk2011}, trap assisted tunneling \cite{Long2012,Puglisi2015,Yu2011,Bersuker2011}, Schottky emission \cite{Lee2012,Syu2013,Khaldi2014}, space charge limited current \cite{Lee2010}, and variable range and nearest neighbor hopping \cite{Fang2015,Sekar2014} have been proposed for similar HfO$_2$ RRAM structures. Here, we tried several of the latter forms, all working well enough and offering no preferences in I-V modeling. At the end, for the sake of its novelty to the community, we chose to use with this publication the phonon assisted hopping through the optimum hopping chains \cite{Shklovskii1976,Levin1988} (see Fig. \ref{Fig:hop}). Its essence is that the electrons utilize the optimum hopping paths combining not too long inter-center tunneling distance and not too large number of hoping centers involved.  The conductivity of such paths vs. electric potential ($V$) and temperature ($T$) is given by,
\begin{eqnarray}
  \sigma_{cf} &=& \sigma_{0f}\exp\left(-\frac{E_{Gf}}{kT}\right) \exp\left(\sqrt{\frac{eV}{kT}}\right), \\
  \sigma_{cg} &=& \sigma_{0g}\exp\left(-\frac{E_{Gg}}{kT}\right) \exp\left(\sqrt{\frac{eV}{kT}}\right),
\end{eqnarray}
where, subscript $f$ and $g$ pertains to filament and gap, $E_{G}$ is the energy difference between the Fermi-level and mobility edge. Note that for this numerical modeling, $E_{Gf}$ and $E_{Gg}$ were not explicitly used, instead the deformation potential components are introduced below.

{\bf Thermal Conduction:} Widemann-Franz-Lorenz law was applied to define thermal conductivity of the metal electrodes and filament similar to Ref. \cite{Niraula22017}. That law was applied as well to the gap thermal transport where the electric conductivity is higher than that of typical insulator; a similar approach was used earlier for GeSbTe layer in modeling phase change memory device. \cite{Kencke2007}

{\bf Ballistic Heat Transport:} One anomaly discovered in this modeling was `unphysically' high (4500 K) temperature of the dielectric layer, which turns out to be well above melting of most of the transition metal oxide \cite{Yaws2015} (e.g. HfO$_2$-2274$^\circ$C, TiO$_2$-1843$^\circ$C, Ta$_2$O$_5$-1784$^\circ$C etc). We then found that similar temperature had been reported for the TaO$_x$ device model in Ref. \cite{Li2017} where the Poole-Frenkel conduction mechanism was employed, however it remained unrecognized as a red flag there.

We have verified that such an `unphysical' temperature is due to the assumed diffusive heat transport. The problem is solved by noticing that the electrons can travel through very thin gap dielectrics ballistically, and each electron carries its corresponding portion of Joule heat away from that gap. It has been observed indeed \cite{Tomita1997} that the energy relaxation length of the electron in amorphous SiO$_2$ is on the order of 10 nm; similar lengths must correspond to HfO$_2$ and other non-crystalline materials. Since, the energy relaxation length is greater than the gap width, most of the energy is dissipated away from the gap. To account for the ballistic transport we introduced the dimensionless corrective coefficient, $\kappa_\textrm{eff}$ defined as the ratio of energy relaxation length in the ballistic heat transfer ($\sim 10$ nm) to that of diffusive heat transfer ($\sim 1$ nm). With that coefficient, the gap thermal conductivity is given by $\kappa_\textrm{gap} = \kappa_\textrm{eff}\kappa_\textrm{diff}$ where $\kappa_\textrm{diff}$ is the diffusive transport value. That procedure allowed us to `cool down' the gap below 1500 K.

\subsubsection{{\bf Switching Condition 1: $V_D = V_T$}}\label{Sec:SC1}
\begin{figure}
  \centering
  \includegraphics[width=0.4\textwidth]{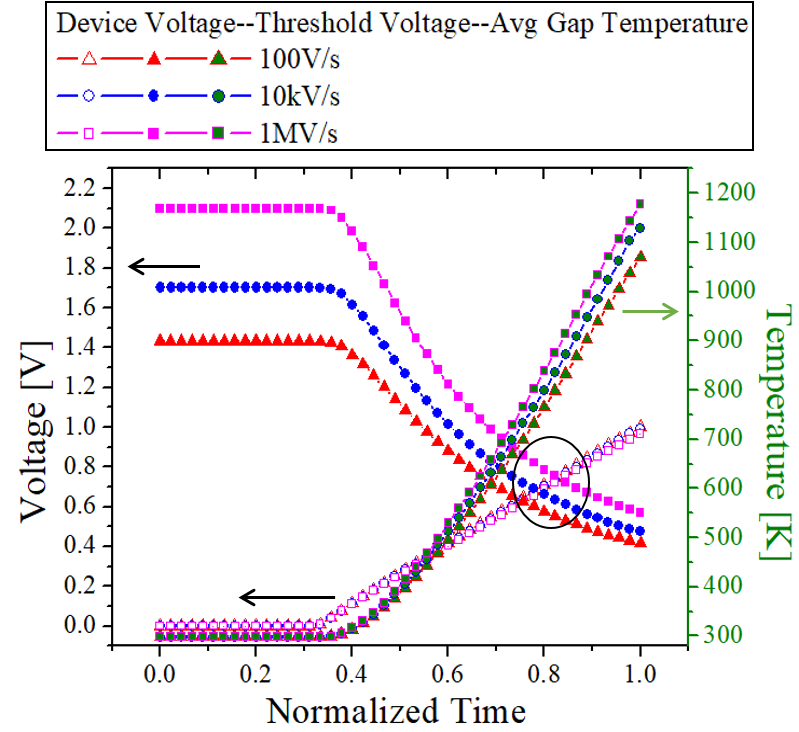}
  \caption{A plot showing device voltages, threshold voltages, and average gap temperatures for 3 different ramp-rates corresponding to half of a pulse. The normalised time is obtained by dividing the time with half of the pulse time. The circle shows the intersection of device voltage and threshold voltage which is the switching condition 2.}\label{Fig:VT}
\end{figure}
Having started with OFF module the program switches to SET module when the input voltage exceeds the threshold value initiating the filament nucleation. The threshold voltage is given by, \cite{Karpov2017}
\begin{equation}\label{eq:VT}
V_T \cong \tilde{V}\left[\ln\left(\frac{\tilde{V}}{\lambda \tau_{0}}\right)\right]^{-1}\mathrm{where},\tilde{V}=\frac{hW_0}{kT}\sqrt{\frac{3\pi^2\alpha^3\Lambda W_0}{32\epsilon r_c^3}},
\end{equation}
where, $h$ is the filament height, $W_0$ is the nucleation barrier, $T$ is the temperature, $\alpha$ is the ratio of minimum filament radius and critical nucleus radius ($r_c$), $\Lambda$ is a multiplier logarithmically dependent on the embryo aspect ratio, $\lambda$ is the voltage ramp-rate, $\tau_0$ is the characteristic atomic-vibration time in solid, and $\epsilon$ is the dielectric permittivity. A subtlety is that the temperature and device voltage (not equal to that of source) depend on the voltage ramp-rate; hence, the entire device operation is involved with the threshold voltage definition.

The threshold voltages for 3 different ramp-rates are illustrated in Fig. \ref{Fig:VT} where we have used the values of parameters listed in Table. \ref{tab:parav}. Note that $V_T$ decreases with temperature which, in turns, increases with the source voltage. Thus the switching condition, $V_D=V_T$, translates into the intersection point of $V_T$ and $V_D$ when plotted against time as presented in Fig. \ref{Fig:VT}.

\begin{figure*}
  \centering
  \includegraphics[width=0.33\textwidth]{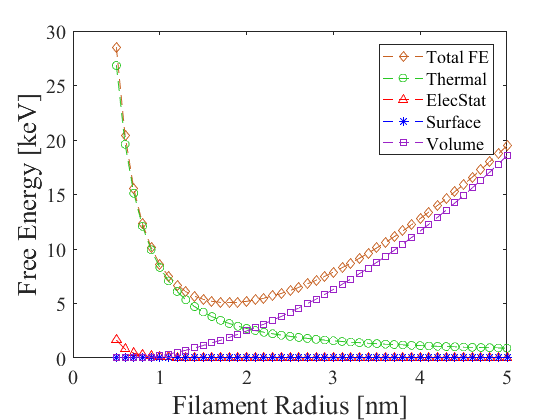}\includegraphics[width=0.33\textwidth]{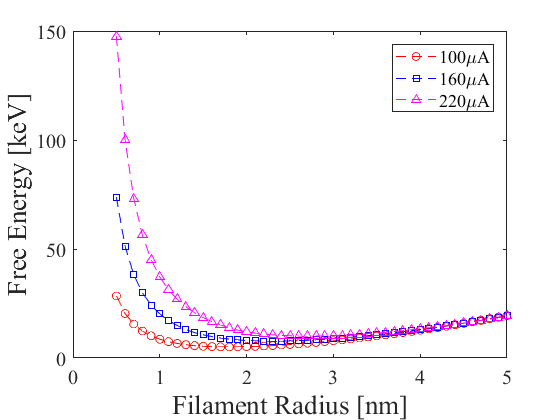}
  \includegraphics[width=0.33\textwidth]{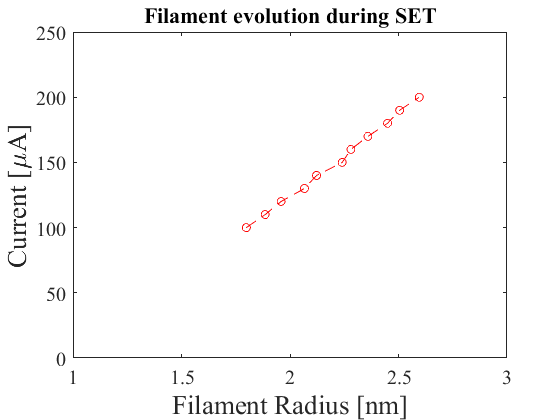}
  \includegraphics[width=0.33\textwidth]{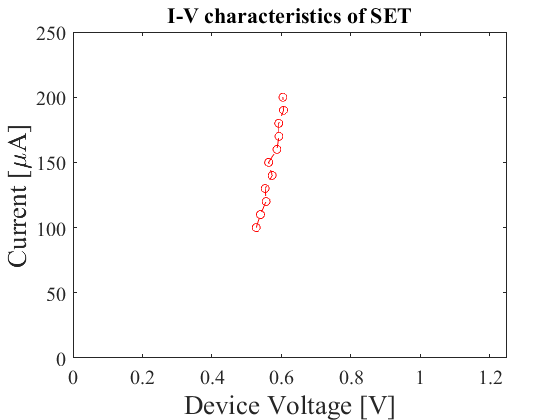}\includegraphics[width=0.33\textwidth]{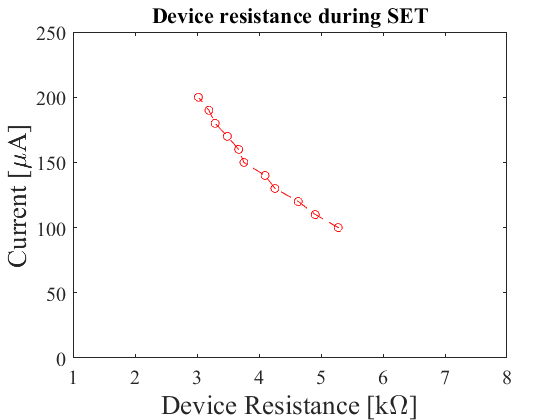}
  \includegraphics[width=0.33\textwidth]{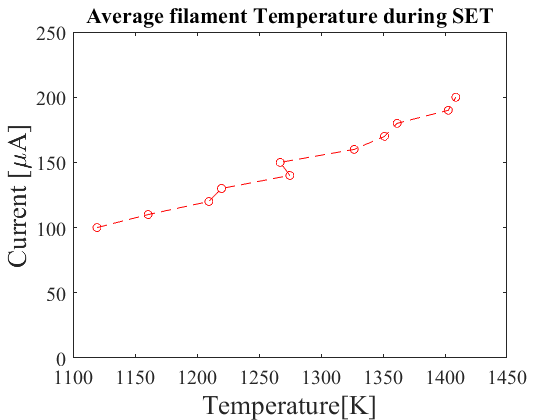}
  \caption{Different simulated results from SET module. Top left: total and partial free energy contribution for 100$\mu A$ current source. Top middle: total free energy for 3 different current source. Top right: stable filament radius as a function of source current. Bottom left: the vertical device I-V corresponding to the stable filament radius. Bottom middle: device resistance during SET process. Bottom Right: average filament temperature during SET process.}\label{Fig:SETFE}
\end{figure*}
\subsubsection{{\bf Module 2: SET}}\label{Sec:M2}
The SET module starts with the device voltage exceeding $V_T$. The device voltage snaps back while the current abruptly increases following the I-V domain A-B, after which the filament grows radially following the ascending power source voltage. The evolution of its radius can be mapped onto the power source voltage by applying the standard thermodynamics principle that the filament radius is determined by the minimization of free energy, which yields the corresponding stable radius.

Assuming the electrodes under certain boundary conditions, we consider the free energy of filament and the dielectric layer only. It consists of
the thermal energy (due to Joule heating), electrostatic energy, and chemical energy (surface and volume), i. e.
\begin{equation}\label{eq:FE}
F = \int \rho C_p \delta T \mathrm{d} x^3 + \frac{1}{2}\int \epsilon |\boldsymbol{E}|^2 \mathrm{d} x^3 + 2 \pi r h \sigma + \pi r^2 h \delta \mu _1,
\end{equation}
where, $\rho$ is the mass density, $C_p$ is the specific heat, $\delta T$ is the change in temperature, $\epsilon$ is the dielectric permittivity, $\boldsymbol{E}$ is the electric field, $r$ is the filament radius, $h$ is its length, $\sigma$ is the interfacial energy, and $\delta \mu _1$ is the difference in the chemical potential between uc-phase and i-phase. Fig. \ref{Fig:SETFE} (top left) shows the free energy components. Practically, the thermal and volume energy contributions are dominant here.

The data in Fig. \ref{Fig:SETFE} (top left) are obtained for the current source regime. To justify that regime, we consider a circuit with two resistors, $R_1$ and $R_2$ in series. When $R_1 \ll R_2$, all the voltage drops across the bigger resistor, i. e. changing $R_1$ does not affect the current. The latter consideration applies to the SET process with $R_1$ understood as the differential resistance ($|dV/dI|$). Since $|dV/dI| \ll R_L$, SET process belongs in the current source regime.

Numerically, the problem reduces to finding the minimum in free energy with respect to the filament radius for a fixed current value. This gives the stable filament radius for a particular value of current. Minimization is then carried out for the next current value in the range of a user input currents. The details of minimization algorithm and method are given in Appendices \ref{Ap:SETMin} and \ref{Ap:MinMethod}. Figure \ref{Fig:SETFE} (top middle) shows the free energy for 3 different current values. Figure \ref{Fig:SETFE} (top right) shows the stable radius for the different current values. The device current and voltage for each stable radius is then extracted which reproduces the vertical SET I-V curve as shown in Fig. \ref{Fig:SETFE} (bottom left). Additionally, Fig. \ref{Fig:SETFE} (bottom middle) and \ref{Fig:SETFE} (bottom right) shows the device resistance and average filament temperature during SET process.

\begin{figure*}
  \centering
  \includegraphics[width=0.33\textwidth]{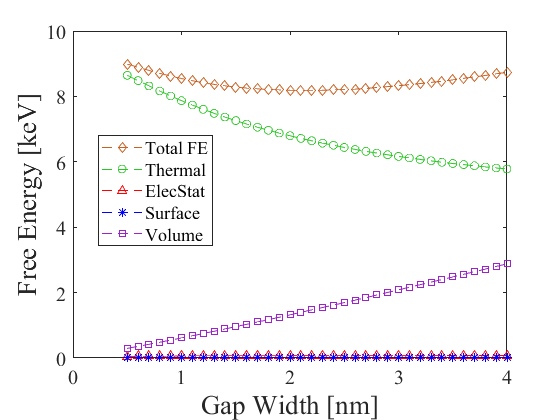}\includegraphics[width=0.33\textwidth]{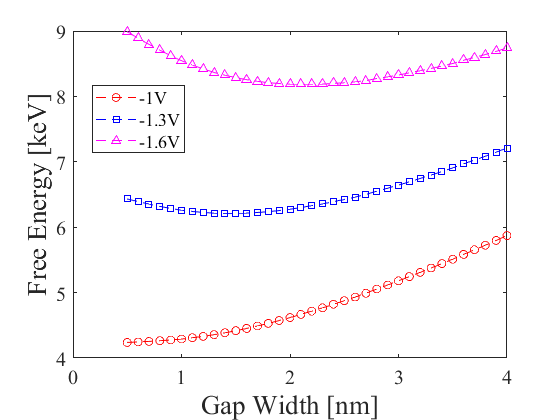}
  \includegraphics[width=0.33\textwidth]{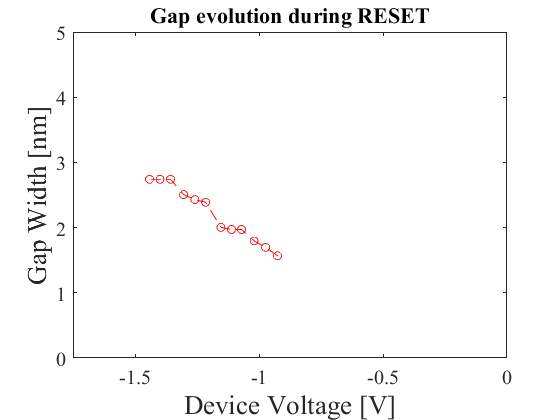}
  \includegraphics[width=0.33\textwidth]{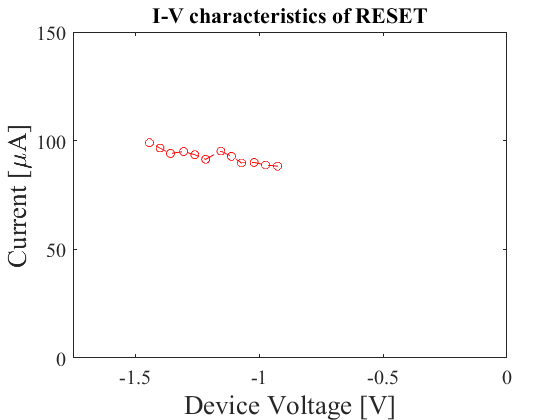}\includegraphics[width=0.33\textwidth]{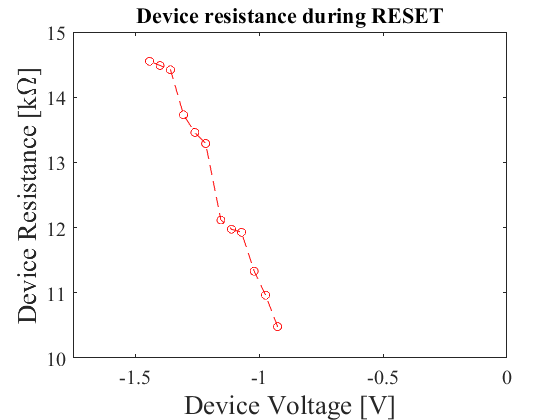}
  \includegraphics[width=0.33\textwidth]{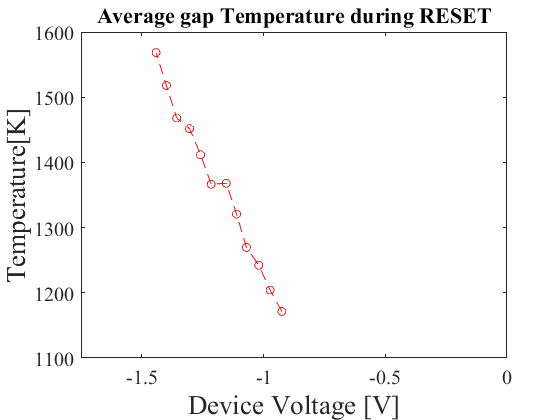}
  \caption{Different simulated results from RESET module. Top left: total and partial free energy contribution for -1.6 V voltage source. Top middle: total free energy for 3 different voltage source. Top right: stable gap width as a function of source voltage. Bottom left: the horizontal device I-V corresponding to the stable gap width. Bottom Middle: device resistance during RESET process. Bottom right: average gap temperature during RESET process.}\label{Fig:RESETFE}
\end{figure*}

\subsubsection{{\bf Switching Condition 2: $V_S = V_\textrm{amp(+ve)}$}}\label{Sec:SC2}
In the SET module, the minimization is carried out until $V_S = V_\textrm{amp(+ve)}$, where $V_S$ is the source voltage and $V_\textrm{amp(+ve)}$ is the amplitude of the positive input voltage pulse, after which the program switches to ON module. In our case $V_S = V_D +IR_L$ is limited by the input pulse amplitude after which the SET process ceases.

In the case of compliance current, this switching condition is more natural; the user input current range during SET process is limited to the compliance current value.

In the former case, the set current $I_\textrm{SET}$ corresponds to the $V_\textrm{amp(+ve)}$ and in the latter case, $I_\textrm{SET}$ equals the compliance current.

\subsubsection{{\bf Module 3 and 4: ON}}\label{Sec:M34}
The ON module begins with the descending part of input pulse where the filament temperature decreases. The filament ``freezes" with its radius fixed throughout the ON state operation. (The radius of fully grown filament is given by the minimum of free energy corresponding to $I_\textrm{SET}$.) The positive voltage regime is completed when the pulse reaches zero and then the simulation switches to the ascending negative voltage regime. The negative voltage regime begins with ON module and switches to the RESET module at the RESET voltage defined as $-V_{\rm SET}$. The ON module produces the I-V domain C-0-D.

\subsubsection{{\bf Switching Condition 3: $T_\textrm{fil} = T_\textrm{I,SET}$}}\label{Sec:SC3}
Since the filament remains intact in ON state, so does its resistance. Thus, when the electric power in device equals to the power at $I_{SET}$ i.e. $I^2R_D=I_{SET}^2R_D$, the filament temperature equals to that at $I_{SET}$, i.e. $T_\textrm{fil}= T_\textrm{I,SET}$. The filament then unfreezes and breaks driven by the unfavorable polarization of the insulating host. That switches the program to the RESET module. The underlying logic is consistent with the observation that RESET current and voltage are equal to those of SET.

\subsubsection{{\bf Module 5: RESET}}\label{Sec:M5}
The RESET process starts with the filament rupture that sends the device back to OFF state. The device voltage snaps forward while the current abruptly decreases following the I-V domain D-E. The gap width grows following the ascending power source voltage. Similar to SET, the gap width evolution is mapped to the source voltage value by applying the thermodynamics and is determined by the minimum in free energy given by,
\begin{equation}\label{eq:FE1}
F = \int \rho C_p \delta T \mathrm{d} x^3 + \frac{1}{2}\int \epsilon |\boldsymbol{E}|^2 \mathrm{d} x^3 + 2 \pi r l \sigma + \pi r^2 l \delta \mu_2,
\end{equation}
where, $l$ is the gap width, and $\delta \mu_2$ is the difference in the chemical potential between uc-phase and mc-phase. Fig. \ref{Fig:RESETFE} (top left) shows the total free energy and its partial components during RESET. Again, the thermal and volume energies dominate.

The free energy in Fig. \ref{Fig:RESETFE} (top left) corresponds to the voltage source regime. Similar to SET process, the device resistance is represented by $|dV/dI| \gg R_L$, i. e. RESET belongs in the voltage source regime. Thus, the problem reduces to finding the free energy  minimum vs. the gap width for a fixed source voltage. The details of the minimization algorithm used are given in Appendices \ref{Ap:RESETMin} and \ref{Ap:MinMethod}. Figure \ref{Fig:RESETFE} (top middle) shows the free energy for 3 different source voltage values. Figure \ref{Fig:RESETFE} (top right) shows the stable gap width for different source voltages. The device current and voltage for each gap width is then extracted reproducing the horizontal RESET I-V curve as shown in Fig. \ref{Fig:RESETFE} (bottom left). Additionally, Fig. \ref{Fig:RESETFE} (bottom middle) and \ref{Fig:RESETFE} (bottom right) shows the device resistance and average gap temperature during RESET.

\subsubsection{{\bf Switching Condition 4: $V_S = V_\textrm{amp(-ve)}$}}\label{Sec:SC4}
When the source voltage equals the negative input voltage pulse amplitude i.e. $V_S = V_\textrm{amp(-ve)}$, the program switches from RESET module to OFF module. The device voltage corresponding to $V_\textrm{amp(-ve)}$ is called $V_\textrm{STOP}$ corresponding to point F in Fig. \ref{Fig:RRAM_IV} (right).

\subsubsection{{\bf Module 6: OFF}}\label{Sec:M6}
The OFF module begins with descending input pulse, during which the gap stabilises. The gap width is determined by the minimum of free energy at $V_\textrm{amp(-ve)}$. The OFF module ends when the voltage reaches zero and yields the I-V domain F-0.\\

\begin{figure}
  \centering
  \includegraphics[width=0.45\textwidth]{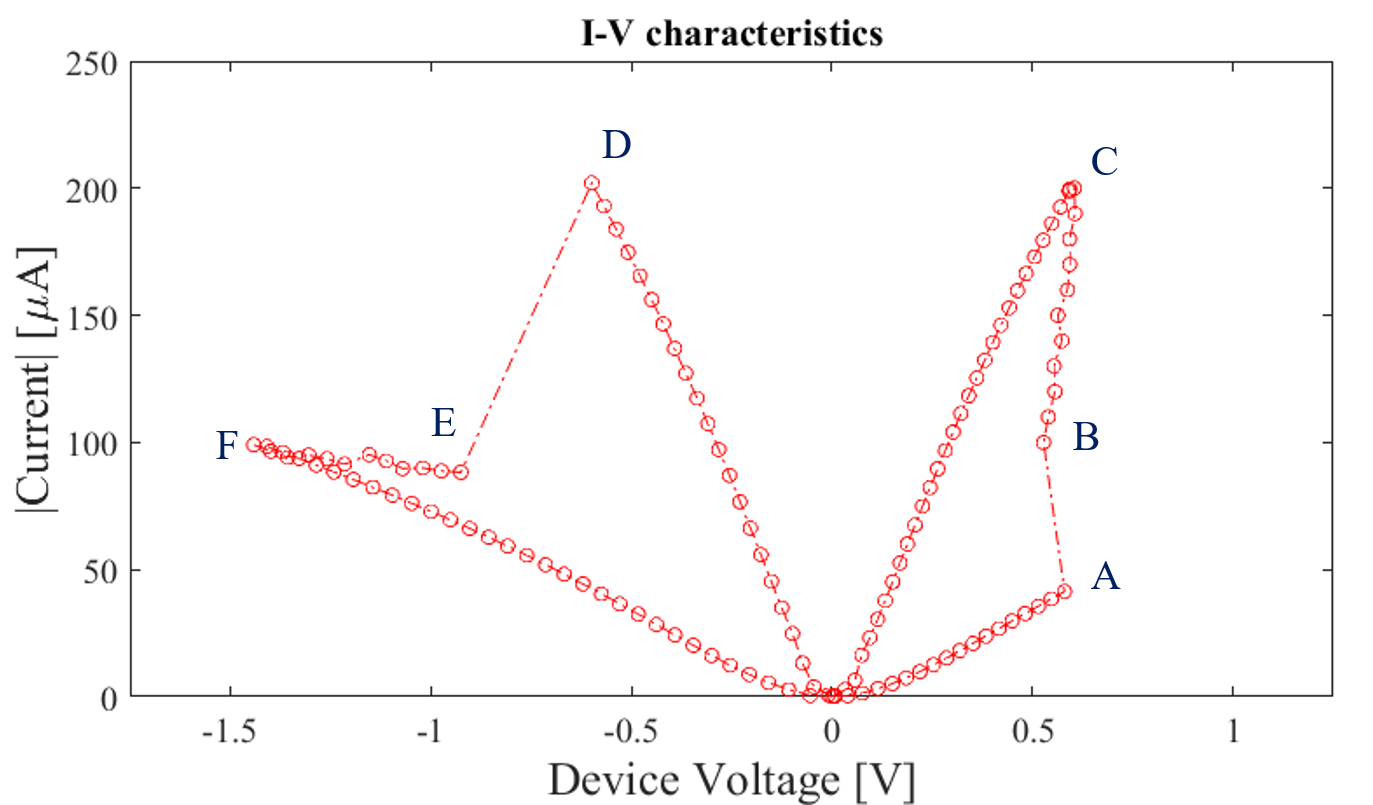}
  \caption{Simulated I-V for 100 V/s ramp-rate.}\label{Fig:IV100}
\end{figure}
This completes one full switching cycle. The corresponding I-V is presented in Fig. \ref{Fig:IV100} for 100 V/s ramp-rate. The ramp-rate dependence and cycle-to-cycle variations are due to the disorder effects described next.

\section{Variability of RRAM parameters: effects of disorder}\label{SEC:Disorder}
The material non-crystallinity is one of the most common features of filamentary RRAM structures. In a non-crystalline material, some atoms or groups of atoms retain a certain mobility being able to move between two equilibrium positions. That feature is described in terms of DWP \cite{Frenkel1955,Anderson1972,Galperin1989,Karpov2007}. (A simplistic structure model of DWP in three-atomic molecule is outlined in Appendix \ref{apen:DWP} to facilitate the intuitive grasp.) The significance of DWP here is that, depending on their local configurations, they will exert this or other force on the system. An obvious example is a DWP caused deformation that, through the deformation potential, translates into shifts of mobility edges making exponentially strong effects on resistivity.

\begin{figure}
  \centering
  \includegraphics[width=0.3\textwidth]{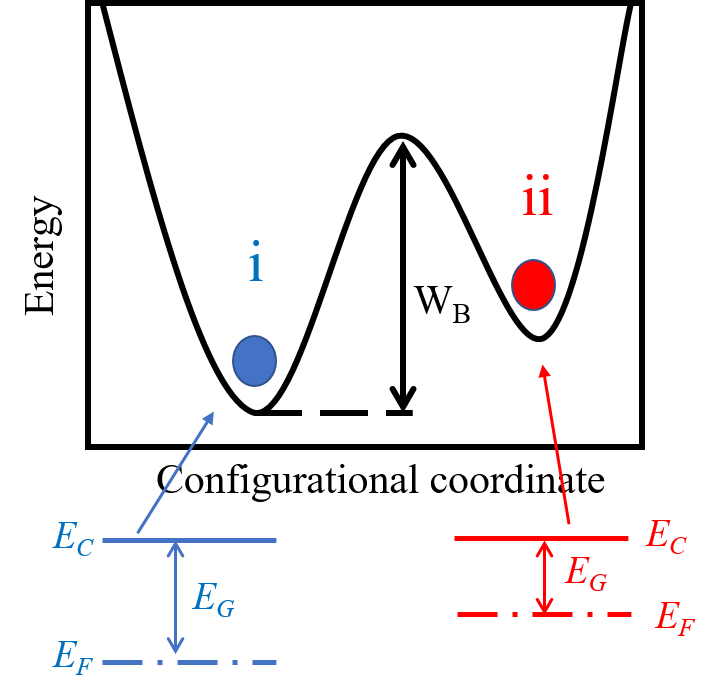}
  \caption{Schematic of a double well potential representing two configuration separated by a barrier height $W_B$. Each configuration is associated with different deformation potential and thus different band gap.}\label{Fig:DefPot}
\end{figure}

Figure \ref{Fig:DefPot} represents the energy diagram of DWP, in which the presence of energy barrier of random height $W_B$ is important. The corresponding relaxation time is given by,
\begin{equation}\label{Eq:transtime}
\tau=\tau_0 \exp \left(\frac{W_B}{kT}\right),\quad\mathrm{i. e.}\quad W_B (\tau) = kT \ln\left(\frac{\tau}{\tau_0}\right).
\end{equation}
The latter relation shows that given a certain relaxation time $\tau$, only DWP with barriers below $W_B(\tau)$ will contribute to the process by changing their configurations.

According to the standard DWP model, the barrier heights are random quantities uniformly distributed in a certain interval ($\Delta W_B$),
\begin{equation}\label{Eq:Barrier}
g(W_B) \approx \frac{1}{\Delta W_B}, \quad \Delta W_B = W_{B,\textrm{max}}-W_{B,\textrm{min}}.
\end{equation}

The corresponding relaxation-time distribution is given by,
\begin{multline}\label{Eq:relaxtime}
\rho(\tau) = g[W_B(\tau)]\left| \frac{dW_B(\tau)}{d\tau}\right|=\frac{kT}{\tau \Delta W_B}\\
 \mathrm{for} \quad \tau_\mathrm{min} < \tau <\tau_\mathrm{max},
\end{multline}
where $\tau_\textrm{min}$ and $\tau_\textrm{max}$ correspond to the minimum and maximum barrier heights respectively.

The gist of DWP application here is as described in the two following subsections.
\subsection{Average rate dependent effects}\label{sec:avrate}
Here, $\tau$ is identified with the voltage pulse length, which is reciprocal of the ramp-rate. Since only DWP with relaxation times shorter than $\tau$ can change their states, the total effect becomes proportional to their fractional concentration,
\begin{equation}\label{Eq:activeDWP}
f(\tau) =\frac{kT}{\Delta W_B}\ln\left(\frac{\tau}{\tau_\mathrm{min}}\right)\quad \mathrm{for}\quad \tau_\mathrm{min} < \tau <\tau_\mathrm{max},
\end{equation}
obtained by integrating Eq. \ref{Eq:relaxtime} from $\tau_\textrm{min}$ to $\tau$. $f(\tau)$ describes the effect of pulse rate on the measured RRAM properties.

The changes in free energy due to phase transformations depend on the differences in chemical potentials presented in Fig. \ref{Fig:phase}. We take into account that the transitions in DWP will change these free energy differentials, which results in,
\begin{eqnarray}
  \delta \mu_1 = \overline{\delta \mu_1}&+& \beta_1 kT \left(\frac{1}{\Delta W_{Buc}}-\frac{1}{\Delta W_{Bi}}\right)\ln\frac{\tau}{\tau_\mathrm{min}}, \label{Eq:mu1}\\
   \delta \mu_2 = \overline{\delta \mu_2}&+& \beta_2 kT \left(\frac{1}{\Delta W_{Buc}}-\frac{1}{\Delta W_{Bmc}}\right)\ln\frac{\tau}{\tau_\mathrm{min}}.\label{Eq:mu2}
\end{eqnarray}
Here $\overline{\delta \mu_1}$ and $\overline{\delta \mu_2}$ represent DWP independent contributions, and $\beta_1$ and $\beta_2$ are the proportionality coefficients to be adjusted in the course of modeling along with $\overline{\delta \mu_1}$ and $\overline{\delta \mu_2}$.

Similarly, the active DWPs also affect the system resistivity \cite{Karpov2007} as mentioned in the above and illustrated in Fig. \ref{Fig:DefPot}. We assume \cite{Karpov2007} that the relative volume change (dilation) is directly proportional to the activated DWP fraction,
\begin{equation}
u(\tau)=u_0 f(\tau).
\end{equation}
Here $u_0$ is the maximum conceivable dilation. Conduction through the filament is an activated process given by the relation, $R_D\propto\exp(E_F/kT)$. The ramp-rate dependence of the resistance comes from the fact that the activation energy is affected by dilation. The deformation dependent activation energy of conduction is given by,
\begin{equation}
E_F \approx \langle E_F \rangle + Du,\quad \mathrm{where,}\quad D\equiv\frac{dE_F}{du},
\end{equation}
where $D$ is the deformation potential whose absolute value of the order of 1 eV is a material parameter. Thus the resistance becomes,
\begin{equation}\label{Eq:Resistance}
R \propto \exp\left[\alpha \ln\left(\frac{\tau}{\tau_{min}}\right)\right],
\end{equation}
where $\alpha = u_0D/\Delta W_B$.

Taking into account Eqs. (\ref{Eq:mu1}), (\ref{Eq:mu2}), and (\ref{Eq:Resistance}), our numerical modeling reproduces ramp-rate dependent I-V characteristics as presented in Fig. \ref{Fig:IV} (top left). The parameters used are listed in Table. \ref{tab:param} and \ref{tab:parav}. We note that the our model is exponentially more sensitive to the rate dependent changes in resistivity than that of chemical potential differentials.

\subsection{Cycle-to-cycle variations}\label{sec:cycle}
The distributions in Eqs. (\ref{Eq:relaxtime}) and (\ref{Eq:activeDWP}) imply large enough systems where the statistics of DWP does not depend on the system size. In nano-size small structures of modern RRAM, the number of DWP may be not large enough to sufficiently present their full spectrum. Using the estimates for the DWP related density of two level system states, \cite{Galperin1989} $10^{32}-10^{33}$ erg$^{-1}$cm$^{-3}$ and assuming the total energy interval of $\sim 0.1$ eV for such systems, one gets an estimate for DWP concentration as $\sim 0.01-0.1$ nm$^{-3}$. That means that 10x10x10 nm$^3$ structure may have total of 10-100 DWPs. Such a scarcity (consistent with the observations of random telegraph noise in nano-sized structures \cite{kogan1996,maestro2016}) makes it possible that different nominally identical structures will have quite different DWP ensembles showing significant variability between their parameters.

Cycle-to-cycle variations find their natural explanation along the same lines because each cycle creates the amorphous portion of the device anew. As a result, every switching cycle under nominally identical conditions yields different microscopic structures. That differences cause variations in DWP barrier heights, chemical potentials, and deformation potentials causing cycle-to-cycle variations in I-V.

\begin{figure*}
  \centering
  \includegraphics[width=0.4\textwidth]{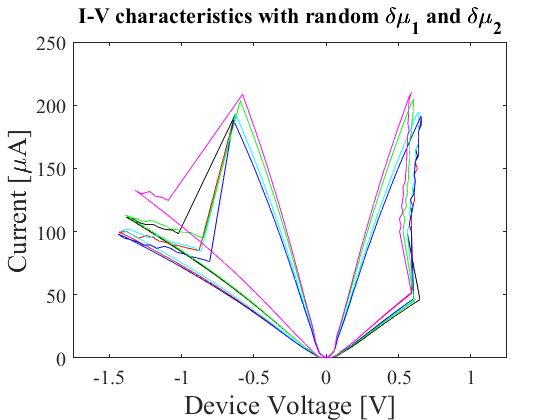}  \includegraphics[width=0.4\textwidth]{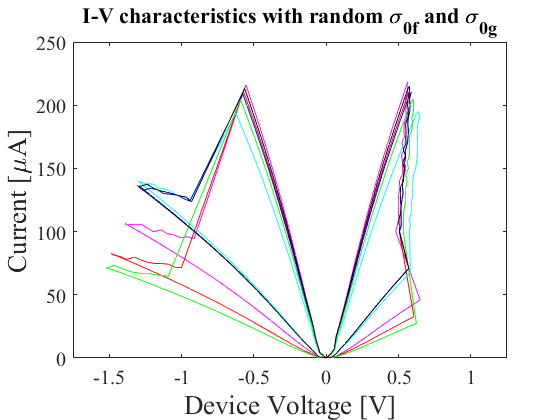}
  \includegraphics[width=0.4\textwidth]{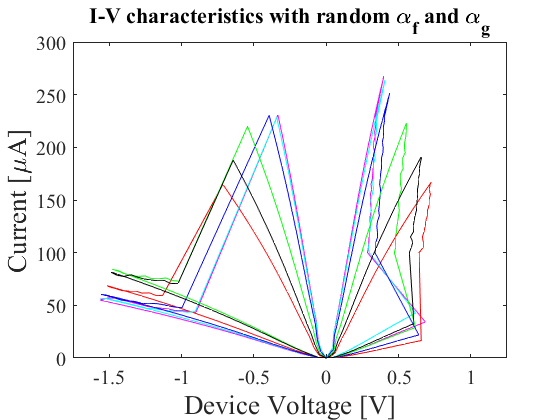}
  \caption{Simulated I-Vs showing cycle-to-cycle variation for 10 cycle with random chemical potential difference (top left), pre-exponential (top right), and pre-logarithmic dependance of conductance (bottom left).} \label{Fig:rand}
\end{figure*}

To account for the latter type of variations, we have applied random number generator to randomly vary the DWP ensembles in a device from one cycle to another thus yielding cycle-to-cycle variations in I-V as presented in Fig. \ref{Fig:rand}. We however ensured that the random selection is made from the `parental' database where all DWP are represented with relative weights described in Eq. (\ref{Eq:Barrier}) and (\ref{Eq:relaxtime}).

For comparison, we have allowed random variations in the DWP deformation potentials in pre-logarithms of the resistivity exponent and in the pre-exponential. The latter variations were incorporated for the following reasons. According to the standard theory of elasticity, \cite{Eshelby1956} a small anisotropic source creates the deformation that decays with distance similar to that of the electric dipole field. Because different DWP can have locations in the filament varying from its central to peripheral regions, their related deformation potential effects will vary; hence, additional source of dispersion. Regarding the pre-exponential in conductivity, we have taken into account that, if the transport includes hopping or tunneling, then the probabilities will appear random in the conductivity pre-exponential.

In Fig. \ref{Fig:rand} (top left) the chemical potential difference ($\overline{\delta \mu}$) of Eqs. (\ref{Eq:mu1}) and (\ref{Eq:mu2}) were randomly varied, in Fig. \ref{Fig:rand} (top right) and in Fig. \ref{Fig:rand} (bottom left) the pre-exponential terms ($\sigma_{0}$) and the pre-logarithmic terms ($\alpha$) of the gap and filament conductivity respectively, were randomly varied. Finally, I-V with all the above expressions simultaneously varied [including the nucleation barrier ($W_0$) of Eq. (\ref{eq:VT})] is presented in Fig. \ref{Fig:rand1} (bottom left). The range of the random functions utilized are presented in Table. \ref{tab:parav}. At this time, our conclusion is that any and all of these types of variations can account for the observed cycle-to-cycle variability. More modeling and experimental research is called upon to identify the most significant of the three sources.

\section{Conclusions}
Several achievements of this work are listed below.\\
(1) We have developed a physics based numerical model of RRAM device that closely reproduces I-V characteristics of a real device uniquely including voltage ramp-rate and cycle-to-cycle variations.\\
(2) The non-trivial features of our model lie in the overlap of the device smallness and its non-crystallinity.\\
(3) The ballistic nature of the electron transport through an amorphous dielectric gap is essential as the only mean to avoid the modeling prediction of unrealistically high device temperatures.\\
(4) The effects of double well atomic potential characteristic of amorphous structures lead to strong variations of parameters between nominally identical RRAM structures. \\
(5) The thermodynamic description (as a great simplification of the generally required kinetic approach) is shown to be sufficient for modeling the nano-size small modern devices.\\
(6) The detailed step-by-step instructions of our modeling and the list of all the parameters, equations, and boundary conditions used are presented.

\section{Acknowledgement}
This work was supported in part by the Semiconductor Research Corporation (SRC) under Contract No. 2016LM-2654.
\appendix
\section{Device Model}\label{Ap:DevModel}
\begin{figure}[!ht]
  \centering
  \includegraphics[width=0.4\textwidth]{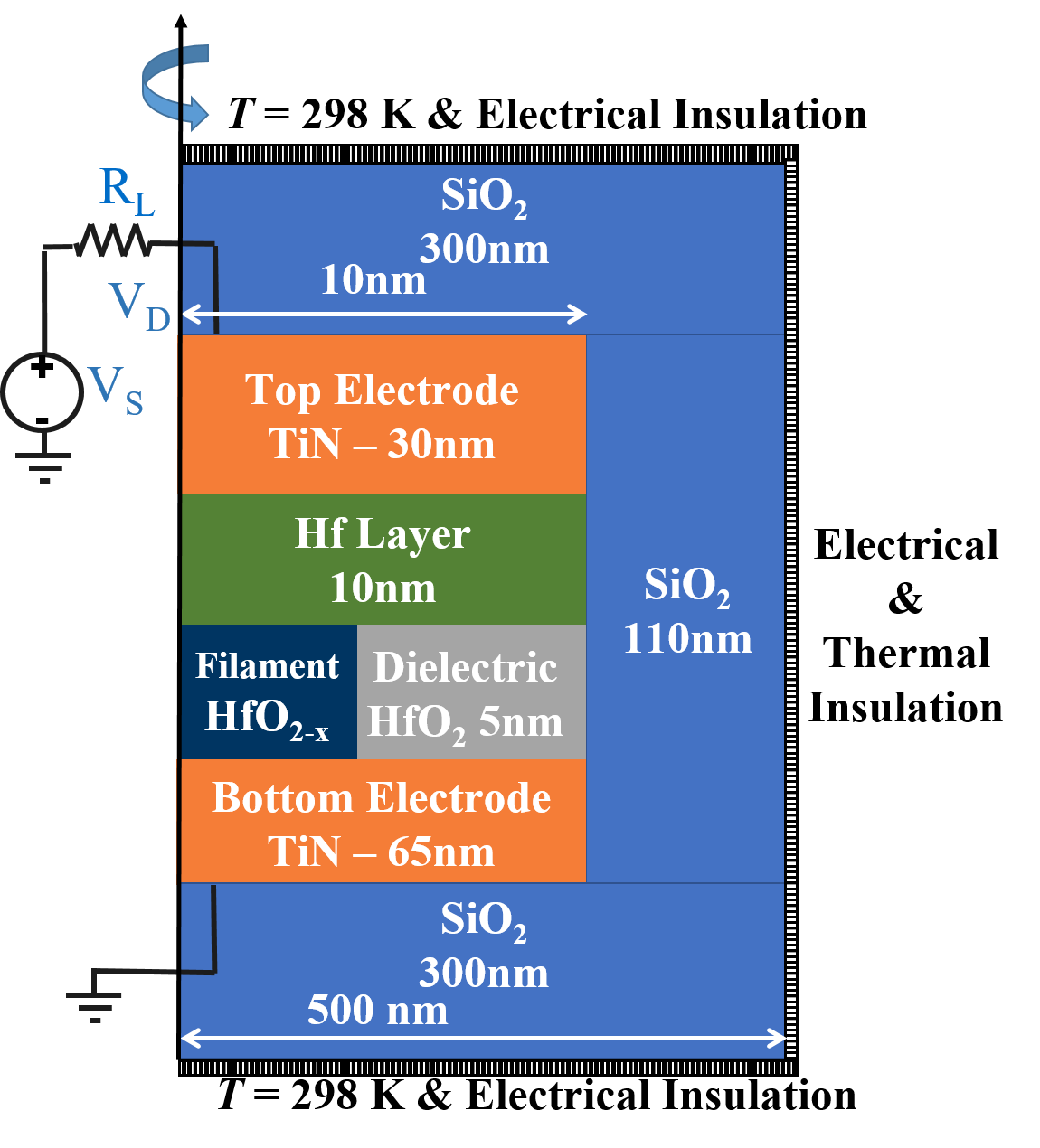}
  \includegraphics[width=0.4\textwidth]{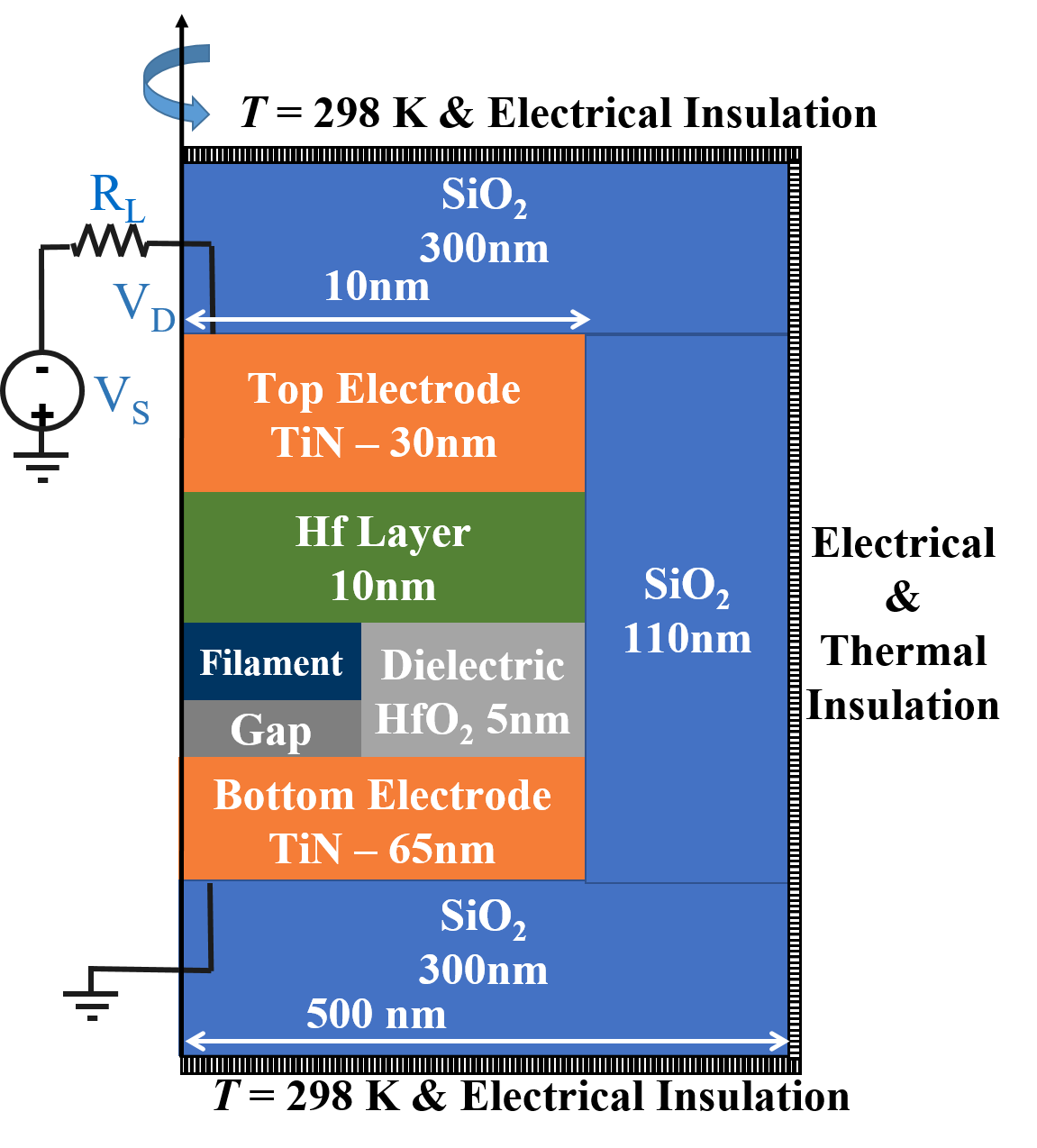}
  \caption{Schematic of multi-layered RRAM cross-section. Top and bottom figures correspond to model used to emulate SET/ON and RESET/OFF process respectively. The figures include geometric parameters and material names of the device’s various layers, and the boundary conditions. Note: Figure not drawn to scale.}\label{fig:dev2018}
\end{figure}

We modeled TiN/Hf/HfO$_2$/TiN multi-layered RRAM device having a cross-sectional area of 20nm $\times$ 20nm as reported in \cite{Fantini2012} and \cite{Fanitini2013}. However, cylindrical geometry of 10 nm radius as in Fig. \ref{fig:dev2018} was considered instead to reduce the 3D to 2D geometry by exploiting the rotational symmetry about the longitudinal axis of the cylindrical shape.

Fig. \ref{fig:dev2018} (top) shows the device model used for ON and SET modules whereas Fig. \ref{fig:dev2018} (bottom) for OFF and RESET modules. Although ON and SET modules utilizes same device model, the filament remains fixed throughout ON whereas it radially evolves in SET. Similarly, the gap remains fixed throughout OFF module whereas it longitudinally evolves in RESET module.

Following the algorithm below, one can simulate RRAM device in COMSOL.

\begin{table*}\footnotesize
\centering
\caption{Material parameters}\label{tab:param}
    \begin{threeparttable}
        \begin{tabular}{Q{2cm} P{6cm} P{2cm} P{2cm} P{1cm} P{2cm}}
        \hline \hline
        Material    & $\sigma_c$[S/m]   & $\kappa$[W/K.m]   & $C_p$[J/kg.K] & $\epsilon_r$\tnote{c}         & $\rho$[kg/m$^3$]\\ \hline
        SiO$_2$     & 10$^{-9}$        & 1.38             & 703           & 3.9                          & 2.2$\cdot10^3$\\
        TiN         & Exp. $\sigma(T)$\tnote{a} &$\sigma_c(T)TL$\tnote{d}       & 545.33        & -$\infty$\tnote{f}            & 5.22$\cdot10^3$ \\
        Hf          & Exp. $\sigma(T)$\tnote{b}  &$\sigma_c(T)TL$ \tnote{d}      & 144           & -$\infty$\tnote{f}            & 13.3$\cdot10^3$\\
        HfO$_2$     & 10\tnote{e}       & 0.5               & 120           & 25                            & 10$\cdot10^3$\\
        HfO$_{2-x}$ & $\sigma_{0f}\exp\left(-\alpha_f \ln\left(\frac{\tau}{\tau_{0}}\right)\right) \exp\left(\sqrt{\frac{eV}{kT}} \right)$ &   $\sigma_{c}(T,V)TL$\tnote{d}   & 140\tnote{e}  & -$\infty$\tnote{e}\tnote{ ,f} & 12$\cdot10^3$\tnote{e}\\
        Gap         & $\sigma_{0g}\exp\left(-\alpha_g \ln\left(\frac{\tau}{\tau_{0}}\right)\right) \exp\left(\sqrt{\frac{eV}{kT}}\right)$ &   $\kappa_\textrm{eff}\sigma_{c}(T,V)TL$\tnote{d}                & 120           & 25            & 10$\cdot10^3$ \\
        \hline \hline
        \end{tabular}
        \begin{tablenotes}
             \item[a]E. Langereis et al., {\it J. Appl. Phys.} {\bf 100}, 023534 (2006).
             \item[b]P.D. Desal, et al., {\it J. Phys. Chem. Ref. Data.} {\bf 3}, 1069 (1984).
             \item[c]Relative Permittivity
             \item[d]Wiedemann--Franz--Lorenz law
             \item[e]Assumed value such that it lies in between Hf and HfO$_2$
             \item[f]-10$^6$ was used instead of -$\infty$ for practical purpose

        \end{tablenotes}
    \end{threeparttable}
\end{table*}

\begin{table}[!ht]\footnotesize
\caption{Various parameters}\label{tab:parav}
\begin{threeparttable}
\begin{tabular}{Q{3cm} P{3cm}}
  \hline \hline
  Parameters                                    &   Value \\
  \hline
  \multicolumn{2}{c}{Circuitry}\\
  \hline
  $R_L$                                         &   3.1 k$\textrm{$\Omega$}$ \\
  $V_\textrm{amp(+ve)}$, $V_\textrm{amp(-ve)}$  &   1.25 V, -1.75 V \\
  $\lambda$                                     &   100 V/s, 10 kV/s, 1 MV/s\\
  \hline
  \multicolumn{2}{c}{Electric Conductivity (gap/filament)}\\
  \hline
  $\sigma_{0f}$                                 &   5 kS/m\\
  $\sigma_{0g}$                                 &   3 kS/m\\
  $\alpha_f$                                    &   -0.05 \\
  $\alpha_g$                                    &   0.05 \\
  $\tau$                                        &   $V_\textrm{amp}/\lambda$ \\
  $\tau_0(\tau_\textrm{min})$                   &   0.1 ps\\
  \hline
  \multicolumn{2}{c}{Thermal Conductivity (gap)}\\
  \hline
  $\kappa_\textrm{eff}$                         &   10\\
  \hline
  \multicolumn{2}{c}{Chemical Energy}\\
  \hline
  $\sigma$                                      &   0.01 J/m$^2$ \\
  $\overline{\delta \mu_1}$                     &   10 GJ/m$^3$ \\
  $\overline{\delta \mu_2}$                     &   6.5 GJ/m$^3$ \\
  $\beta_1$                                     &   0.35 GJ/m$^3$ \\
  $\beta_2$                                     &   0.5 GJ/m$^3$ \\
  $\Delta W_{Buc}$                              &   1.0 eV\\
  $\Delta W_{Bi}$                               &   0.1 eV\\
  $\Delta W_{Bmc}$                              &   0.3 eV\\
  \hline
  \multicolumn{2}{c}{Filament Nucleation}\\
  \hline
  $h$                                           &   5 nm\\
  $W_0$                                         &   2.5 eV\\
  $\Lambda$                                     &   6.6\\
  $r_c$                                         &   2.9 nm\\
  $r_\textrm{min}$                              &   0.5 nm\\
  $\alpha$                                      &   $r_\textrm{min}/r_c$\\
  \hline
  \multicolumn{2}{c}{Static Disorder}\\
  \hline
  $\alpha_f$                                    &   rand(-0.07, -0.03)\tnote{g} \\
  $\alpha_g$                                    &   rand(0.03, 0.07)\tnote{g} \\
  $\sigma_{0f}$                                 &   rand(2,8) kS/m \tnote{g}\\
  $\sigma_{0g}$                                 &   rand(1,5) kS/m \tnote{g}\\
  $\overline{\delta \mu_1}$                     &   rand(8.5,11.5) GJ/m$^3$\tnote{g}\\
  $\overline{\delta \mu_2}$                     &   rand(5.5, 7.5) GJ/m$^3$\tnote{g}\\
  $W_0$                                         &   rand(2.4, 2.6) eV\tnote{g}\\
  \hline
  \hline
\end{tabular}
    \begin{tablenotes}
        \item[g] function rand(x,y) produces uniformly distributed random number between x and y
    \end{tablenotes}
\end{threeparttable}
\end{table}
\subsection{COMSOL Algorithm}
Following algorithm pertains to COMSOL version 5.3a.
\begin{enumerate}[nolistsep]
\item Open {\bf Model Wizard}
\item Select {\bf 2D Axisymmetric} as {\bf Space Dimension}
\item Select {\bf AC/DC module} and add {\bf Electric Currents} and {\bf Electrical Circuit} submodules
\item Select {\bf Heat Transfer Module} and add {\bf Heat Transfer in Solids} submodule
\item Select {\bf Done}
\item Create {\bf Geometry} of the device as in Fig. \ref{fig:dev2018}
\item Create {\bf Blank Materials} in the {\bf Materials} node and add material parameters from Table \ref{tab:param}
\item To add the experimental temperature-dependent electric conductivity,
    \begin{itemize}[nolistsep]
        \item from {\bf Definitions} node, select {\bf Functions} then {\bf Interpolation}, and then insert temperature and corresponding conductivity values in the given table
        \item select {\bf Linear} in both {\bf Interpolation} and {\bf Extrapolation} option
    \end{itemize}
\item To add the temperature and voltage dependent hopping conductivity,
    \begin{itemize}[nolistsep]
        \item from {\bf Definitions} node, select {\bf Variables}, add the formula from Table \ref{tab:param}, and then select the corresponding domain.
        \item Note: the argument of the exponential and logarithm function must be unitless
    \end{itemize}
\item Assign the materials to the corresponding domain.
\item In {\bf Electric Currents} submodule,
    \begin{itemize}[nolistsep]
        \item add {\bf Terminal} boundary condition, select the top boundary of the top electrode, and then select {\bf Circuit} as the {\bf Terminal type}
         \item add {\bf Ground} boundary condition and select the bottom boundary of the bottom electrode
        \item Note: {\bf Electric Currents} submodule has four necessary default subnodes
    \end{itemize}
\item In {\bf Electric Circuit} submodule,
    \begin{itemize}[nolistsep]
        \item add {\bf Resistor} and insert the value of load resistance,
        \item add {\bf External I Vs. U} from {\bf External Coupling} and select {\bf Terminal voltage} from {\bf Electric potential} option
        \item add {\bf Voltage Source} for OFF, ON, and RESET modules and {\bf Current Source} for SET module
        \item select {\bf DC-Source} as {\bf Source type} for SET and RESET modules then insert the source current or voltage value
        \item select {\bf Pulse source} as {\bf Source type} for ON and OFF modules and define the pulse length according to the different ramp-rate as listed in Table. \ref{tab:parav}
        \item Note: every component has positive {\bf p} and negative {\bf n} node names under {\bf Node Connections} which enables to position a component in the circuit. For example, {\bf Ground Node} is {\bf 0} by default; to ground the voltage source, insert {\bf 0} in the {\bf n} node name
    \item Note: {\bf Electric Circuit} submodule has one necessary default subnode
    \end{itemize}
\item In {\bf Heat Transfer in Solids} submodule,
    \begin{itemize}[nolistsep]
        \item add {\bf Temperature} boundary condition, select the top boundary of the SiO$_2$ superstrate and bottom boundary of the SiO$_2$ substrate, and choose 298K in the user defined temperature section.
        \item add {\bf Diffusive Surface} boundary condition, select all the inner boundaries, and then choose 298K in the user defined temperature section and 0.9 in the user defined {\bf Surface emissivity} section
        \item Note: {\bf Heat Transfer in Solids} submodule has four necessary default subnodes
    \end{itemize}
\item In {\bf Multiphysics} node, to couple the {\bf Electric Currents} and {\bf Heat Transfer in Solids} submodule,
    \begin{itemize}[nolistsep]
        \item select all the domains and boundaries in {\bf Electromagnetic Heating} sub-node
        \item select {\bf Heat Transfer in Solid} as {\bf Source} and {\bf Electric Currents} as {\bf Destination} in {\bf Temperature} sub-node
    \end{itemize}
\item Create {\bf Mesh}
    \begin{itemize}[nolistsep]
        \item either automatic {\bf Physics-controlled mesh} or manual {\bf User-controlled mesh} can be selected
        \item {\bf Free Triangular} meshes of different sizes were manually defined for our simulation--to define the mesh size, add {\bf Size} in {\bf Free Triangular} mesh, then use either the {\bf Predefined} or the {\bf Custom} option
        \item select {\bf Build All}
    \end{itemize}
\item Select {\bf Study} type
    \begin{itemize}[nolistsep]
        \item select {\bf Time Dependent} study for ON and OFF modules and then add {\bf Times} corresponding to the pulse lengths
        \item select {\bf Stationary} study for SET and RESET modules
    \end{itemize}
\item Select {\bf Compute}
\item Obtain results in desired form from the {\bf Results} node
\end{enumerate}
\subsection{Differential Equations}
For ON and OFF modules, COMSOL solves the following time-dependent equations,
\begin{enumerate}
  \item {\bf Electric Currents} module:
    \begin{equation}\label{eq:ECtdep}
    \boldsymbol{\nabla}.\boldsymbol{J}=0,\quad\boldsymbol{J}=\sigma_c\boldsymbol{E}+\epsilon\frac{\partial \boldsymbol{E}}{\partial t},\quad\boldsymbol{E}=-\nabla V.
    \end{equation}
  \item {\bf Heat Transfer in Solids} module:
  \begin{equation}\label{eq:HTtdep}
  \rho C_p\frac{\partial T}{\partial t}-\nabla.(\kappa \nabla T)=Q_s.
  \end{equation}
  \item {\bf Multiphysics} module:
  \begin{equation}\label{eq:Multdep}
  Q_s = \boldsymbol{J}.\boldsymbol{E}.
  \end{equation}
\end{enumerate}
For SET and RESET module, COMSOL solves the following stationary equations,
\begin{enumerate}
  \item {\bf Electric Currents} module:
    \begin{equation}\label{eq:ECtind}
    \boldsymbol{\nabla}.\boldsymbol{J}=0,\quad\boldsymbol{J}=\sigma_c\boldsymbol{E},\quad\boldsymbol{E}=-\nabla V.
    \end{equation}
  \item {\bf Heat Transfer in Solids} module:
  \begin{equation}\label{eq:HTtind}
   -\nabla.(\kappa \nabla T)=Q_s.
  \end{equation}
  \item {\bf Multiphysics} module:
  \begin{equation}\label{eq:Multind}
  Q_s = \boldsymbol{J}.\boldsymbol{E}.
  \end{equation}
\end{enumerate}

In the above equations, $\boldsymbol {J}$ is the current density, $\sigma_c$ is the electric conductivity, $\boldsymbol{E}$ is the electric field, $\kappa$ is the thermal conductivity, and $Q_s$ is the heat source. Equations in Eq. (\ref{eq:ECtdep}) and (\ref{eq:ECtind}) are the current conservation law, Ohms law, and the relation between electric field and electric potential due to Maxwell law respectively. Equations (\ref{eq:HTtdep}) and (\ref{eq:HTtind})  are the Fourier heat law where the heat source is represented by the Joule heat terms in Eqs. (\ref{eq:Multdep}) and (\ref{eq:Multind}). The Multiphysics module couples the Electric Currents and Heat Transfer in Solids modules to define the electromagnetic heat source.

\subsection{Boundary Conditions}
The required boundary conditions are listed below.
\begin{enumerate}[nolistsep]
\item {\bf Electric Insulation} ($\boldsymbol{n}.\boldsymbol{J} = 0$)\\
All three boundaries of the SiO$_2$ domain are electrically insulated. Here {\bf n} is the unit vector normal to the surface boundary.
\item {\bf Ground} ($V$ = 0)\\
Interface common to the bottom electrode and SiO$_2$ substrate is grounded.
\item {\bf Terminal} ($V = V_D$ or $I=I_S$)\\
Interface common to the top electrode and SiO$_2$ superstrate is connected to the power-source and load resistor in series. {\bf Terminal} handles both voltage source and current source.
\item {\bf Temperature} ($T$ = 298 K)\\
Free surfaces of both SiO$_2$ layers are placed at room temperature, assuming they are in contact with a larger body which acts as a heat sink and maintains room temperature. Also, all three boundaries of the air domain are at room temperature.
\item {\bf Diffuse Surface} ($-\boldsymbol{n}.\boldsymbol{q} = \sigma_B (T_\textrm{amb}^4 - T^4)$)\\
All the interface of the device loses heat through radiation governed by Stefan-Boltzmann law. Here q is the power radiated per
surface area, $\sigma_B$ is the Stefan-Boltzmann constant, and $T_\textrm{amb}$ is the ambient temperature (298 K).
\end{enumerate}

\subsection{Parameters}
The geometric parameters are presented in Fig. \ref{fig:dev2018} and coefficients of the differential and free energy equations used for the modeling are listed in Table. \ref{tab:param} and \ref{tab:parav}.

\section{Minimization Algorithm}
This section presents two sets of the minimization algorithms corresponding to SET and RESET process, formulated for a computer program followed by the details of the method applied.
\subsection{SET Process}\label{Ap:SETMin}
\begin{enumerate}[nolistsep]
\item construct a device
\item fix the current and calculate free energy of the device for all filament radii: 0 to device radius\label{SETitem1}
\item find the minimum in the free energy with respect to filament radius ($\partial F/\partial r$) for a fix current\label{SETitem2}
\item redo step \ref{SETitem1} and \ref{SETitem2} for all current values taking discrete steps
\item record the device I-V corresponding to the free energy minimum for all current\label{SETitem3}
\item vertical SET I-V is obtained in step \ref{SETitem3}
\end{enumerate}

\subsection{RESET Process}\label{Ap:RESETMin}
\begin{enumerate}[nolistsep]
\item construct a device
\item fix the source voltage and calculate free energy of the device for all gap width: 0 to insulator thickness\label{RESETitem1}
\item find the minimum in the free energy with respect to gap width ($\partial F/\partial l$) for a fix source voltage\label{RESETitem2}
\item redo step \ref{RESETitem1} and \ref{RESETitem2} for all source voltages taking discrete steps
\item record the device I-V corresponding to the free energy minimum for all source voltages\label{RESETitem3}
\item horizontal RESET I-V is obtained in step \ref{RESETitem3}
\end{enumerate}

\subsection{Minimization Procedure}\label{Ap:MinMethod}
Brent's Method \cite{Press1996} is utilized to find the free energy minimum. The method details are not included in this report and can be found in any relevant standard text. However, one important detail is that the method will yield one of the boundary free energy values if there exist no minima in the range of filament radii/gap width. Thus, the minimization process is divided into two steps: (1) minimum in free energy is located with low resolution, (2) if there exist a minimum, then Brent's method is applied.

For instance, the free energy minimization with respect to filament radius was carried out ranging from 1 nm to device radius with a step ($\Delta r$) of 2 nm, then the minimum is searched, by applying the condition
\begin{multline*}
\mathrm{Is}\quad F(r_i+\Delta r)-F(r_i) > 0 \\
\quad \mathrm{and}\quad F(r_i)-F(r_i-\Delta r)<0 ?
\end{multline*}
The value of $r_i$ that yields true for the above condition is the stable radius ($r_c$), and $F(r_c)$ is the minimum for that resolution. If the condition is not satisfied by any filament radius, then there exists no minima for that particular current, and the program moves to next value of current. When the condition is satisfied, the program applies Brent's method for the filament range, $r_c - \Delta r$ to $r_c +\Delta r$. This yields a more accurate stable radius.

\begin{figure}[t]
  \centering
  \includegraphics[width=0.5\textwidth]{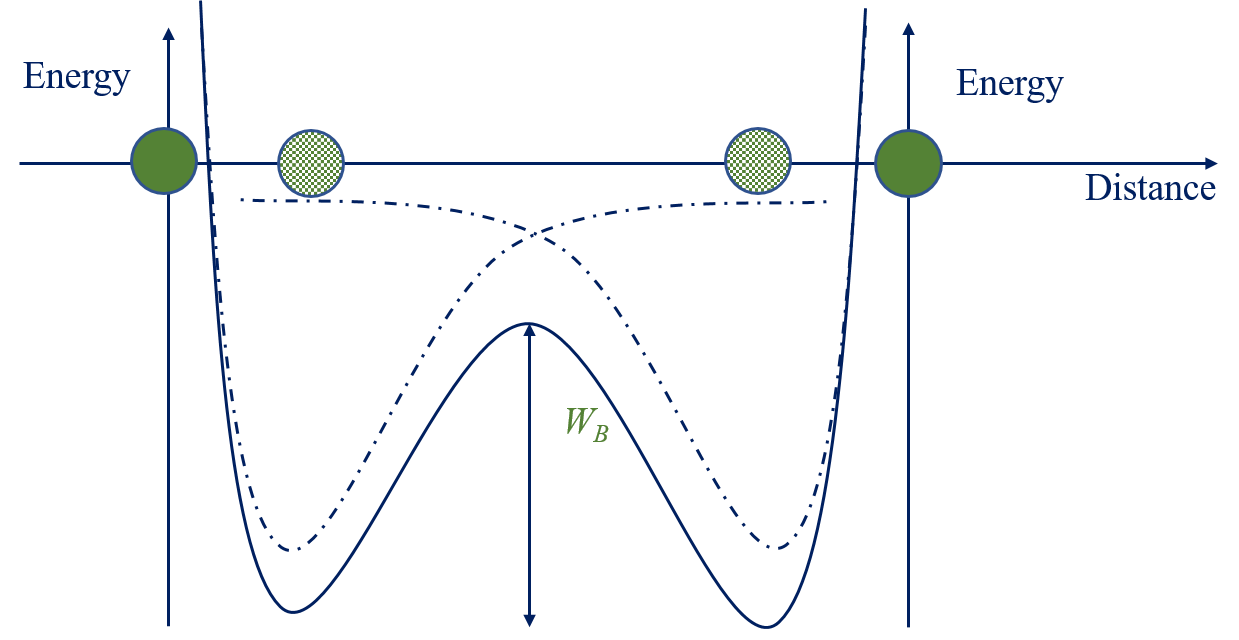}
  \caption{Simplest representation of double well potential due to 2 identical green atoms in a solid. The faded green atom represents a third atom that can exist in either side of the barrier. $W_B$ is the barrier height.}\label{Fig:exDWP}
\end{figure}

\section{A simple model of DWP}\label{apen:DWP}
Consider a three-atomic molecule in Fig. \ref{Fig:exDWP}, which total length is fixed. The atom in the middle moves in the potential obtained by the summation of the two pair potentials shown by the dashed lines. That sum yields a DWP when the total length of the molecule is large enough; however, it represents a single-well potential for short molecules. Because such molecules are randomly stretched in an amorphous structure, the corresponding DWPs are unavoidable. The dispersion of molecule lengths and the fields of other atoms make the DWP parameters random. \cite{Galperin1989}

\end{document}